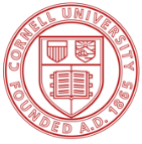
REPORT: Grant 2024-RRF-Arson-cfa36
Cornell Atkinson for Sustainability
**Workshop: The role of EGS in the energy transition at Cornell,** Cornell University, Ithaca, October 23-24, 2024
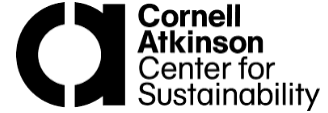

# The role of Enhanced Geothermal Systems in the energy transition at Cornell

# Report of a workshop held at Cornell University, Ithaca, October 23-24, 2024

Chloé Arson, Dominic Balog-Way, Koenraad Beckers, Wayne Bezner-Kerr, Sarah Carson, Stacey Edwards, Patrick Fulton, Michael Gillenwater, Trystan Goetze, Olaf Gustafson, Tony Ingraffea, Terry Jordan, Katherine McComas, Sheila Olmstead, Seth Saltiel, Jeff Tester, Cole Tucker, Marguerite Wells

December 2024

# Table of contents

# Introduction

Various renewable energy technologies are currently being deployed on the Ithaca campus of Cornell University[1]. The development of geothermal infrastructure for direct heat use is still under study. In 2010, Cornell received an Integrative Graduate Education and Research Training (IGERT) grant from the National Science Foundation (NSF) to develop an education program on sustainable subsurface energy engineering and geoscience[2]. In 2022, the Cornell Earth Source Heat (ESH) team championed the drilling of the Cornell University Borehole Observatory (CUBO)[3] with financial support from the Department of Energy (DOE)[4]. The ESH Community Advisory Team (CAT) has communicated regularly with various local stakeholders to educate on Enhanced Geothermal Systems (EGS), share information about CUBO, and enhance public acceptance of deep geothermal energy[5,6]. The next ESH goal is to create a prototype EGS on campus, comprising an injection well and a production well at the vicinity to CUBO. To this aim, a group of engineers from the Cornell energy and sustainability offices and ten Cornell faculty members in Civil and Environmental Engineering, Earth and Atmospheric Sciences, Chemical and Biomolecular Engineering, Communications, and Public Policy submitted a proposal to the DOE[7] in September 2024 to deepen the CUBO well, enhance the instrumentation of the site, and model several scenarios of drilling and stimulation of a geothermal well pair on campus. The group is actively seeking additional funding to support the construction of the prototype geothermal well pair and a subsequent full-scale flow test. To summarize the lessons learnt from recent deep geothermal case studies and plan strategically the research, development, regulation, and communication work required for the implementation of EGS at Cornell University, the same group of engineers and scholars convened a two-day workshop on the Ithaca campus, on October 23-24, 2024. The event was funded by Cornell Atkinson Center for Sustainability[8].

The workshop gathered an inter-disciplinary group of Cornell faculty members who have already collaborated on subsurface energy systems, as well as members of the ESH project, and external stakeholders from the National Renewable Energy Laboratory (NREL), the Greenhouse Gas management institute (GHG), and the Alliance for Clean Energy (ACE) based in New York State. The workshop opened with an overview of Cornell's climate goals and Cornell's energy consumption, production and resources. Then, the group reflected on the concept of public trust in renewable energy, the economic value of reducing carbon emissions, the lessons learnt from the fossil fuel industry for future EGS regulation, and risk communication. Geoscience and engineering results obtained from instrumentation, analysis and modeling at CUBO and at the DOE-funded Frontier Observatory for Research in Geothermal Energy (FORGE) in Utah were then reviewed. Then, a state-of-the-art of the geomechanical and geophysical research landscape was discussed. This was followed by a detailed account of the anticipated design and construction steps of an EGS at Cornell. The workshop concluded with a panel on the environmental impacts of EGS. Appendix 1 provides the full agenda of the workshop, while the list of registered participants is given in Appendix 2.

---

[1] https://sustainablecampus.cornell.edu/news/countdown-net-zero-emissions
[2] IGERT: Training Program in Sustainable Energy Recovery from the Earth - Educational Innovation at the Intersection of Geosciences and Engineering, National Science Foundation (NSF) Integrative Graduate Education and Research Training program (IGERT), grant DGE-0966045, 2010-2017. Lead-PI: J. Tester. Co-PIs: D. Koch, T. Jordan, A. Ingraffea.
[3] https://sustainablecampus.cornell.edu/news/2-mile-borehole-reveal-viability-campuss-geothermal-future
[4] Ground-Truthing: Exploratory Borehole Characterization and Modeling to Expand Techno-Economic Evaluation of Earth Source Heat at Cornell University, U.S. Department of Energy's Office of Energy Efficiency and Renewable Energy (EERE) under the Geothermal Technologies Office, Award DE-EE0009255, 2022. Lead-PI: J. Tester.
[5] https://news.cornell.edu/stories/2021/11/earth-source-heat-open-house-addresses-community-questions
[6] https://www.cornell.edu/video/earth-source-heat-community-forum-january-2020
[7] Demonstration of Eastern US Enhanced Geothermal System (Cornell University), proposal submitted to the Department of Energy, EGS demonstrations. Total budget: $17M including $14.2M from DOE and $2.8M in cost-sharing. Lead PI: B. Bland. Co-PI: C. Arson. Aspiring PI: P. Fulton. Senior advisors: G. Abers, J. Tester.
[8] Workshop: The role of Enhanced Geothermal Systems in the energy transition at Cornell, Cornell Atkinson Center for Sustainability, Rapid Response Fund, Grant 2024-RRF-Arson-cfa36, 2024. PI: C. Arson.

Ten to fifteen additional participants attended parts of the workshop without registering – mostly graduate students and postdoctoral researchers.

The following is a summary of the content of the presentations and discussions that took place during the workshop. The first section focuses on philosophical, sociological, economic, and regulatory questions posed by EGS deployment as a means to mitigate climate change. The second section tackles the scientific and technological research areas associated with EGS. The third section aims to assess the feasibility of developing EGS for heat direct use at Cornell University, based on results and information available to date. The report concludes with a summary of the most salient technological and scientific breakthroughs, and a plan for future technological and academic engagement in EGS projects at Cornell.

## EGS philosophical, sociological, economic, and regulatory questions

**Public Trust and Renewable Energy (Trystan Goetze)**

Local renewable energy developments often face fierce opposition, and yet, according to a 2023 study by the Pew Center[9], two-thirds of Americans prioritize developing alternative energy sources, like wind and solar energy. The apparent social gap is often attributed to ignorance or NIMBY-ism (where NIMBY stands for "Not In My Back Yard")[10]. Assuming ignorance implies that the argument against local renewable energy lacks information on scientific, environmental or economic facts, and establishes the precedence of science over conscience. Attributing the reluctance to renewables to NIMBY-ism reduces opposition to alternative energy to a form of social disengagement. If ignorance and NIMBY-ism were indeed the main causes of local opposition to renewable energy projects, then, it would be possible to provide facts to persuade communities to accept renewables. But often, providing facts is not effective. Research suggests indeed that lack of public trust and misalignment of values are more likely explanations for local opposition to alternative energy developments[10,11]. Rather than persuading people to accept a renewable energy project, it is more effective and ethical to engage in a sustained collaboration to build trust, align values, and coproduce solutions. Trust requires vulnerability (of the truster) and accountability (of the trustee)[12]. If trust requires normative expectations and accountability, then, the truster and trustee need to agree on values. For instance, engineers developing a geothermal project and residents of the targeted geothermal installation ought to share some environmental, economic, sociological or ethical values, otherwise, the residents (trusters) may formulate expectations that are not compatible with the practical reasoning of the engineers (trustees). A misalignment of values threatens trust and may create an unstable and potentially hostile relationship. Pursuing the social license to operate (SLO) is one way of building trust. The SLO is defined as the "implied consent from affected stakeholders towards projects developed by businesses or industries, independent from legal or statutory requirements" (Barich et al. 2022)[13]. The SLO is the strongest when developers and affected communities co-own the project, as illustrated in Figure 1.

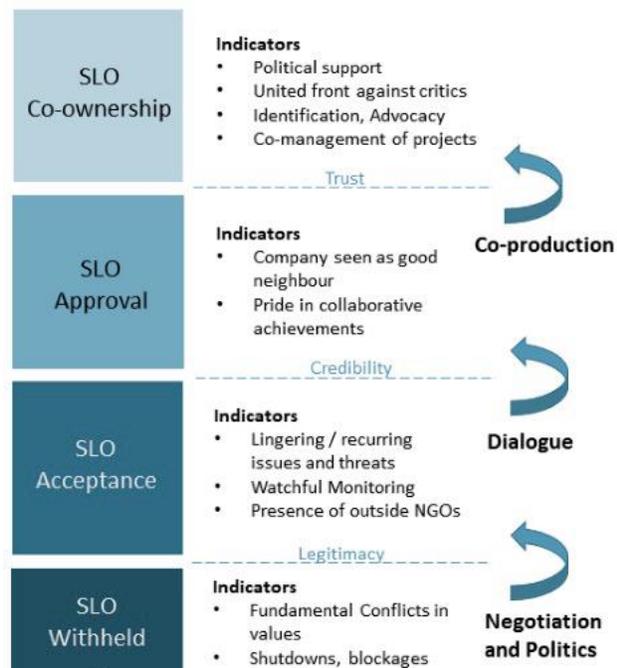

**Figure 1**. Social License to Operate (SLO), from Barich et al. (2022)[13].

---

[9] https://www.pewresearch.org/science/wp-content/uploads/sites/16/2023/06/PS_2023.06.25_climate-energy_REPORT.pdf
[10] Bell, D., T. Gray, and C. Haggett, 2005, The 'Social Gap' in Wind Farm Siting Decisions: Explanations and Policy Responses, Environmental Politics 14.4: 460–77, doi: 10.1080/09644010500175833
[11] Nilson, R.S. and R.C. Stedman, 2023, Reacting to the Rural Burden: Understanding Opposition to Utility-Scale Solar Development in Upstate New York, Rural Sociology 88.2: 578–605, doi: 10.1111/ruso.12486
[12] Walker, M.U., 2006, Moral Repair: Reconstructing Moral Relations after Wrongdoing, Cambridge, UK: Cambridge University Press, doi: 10.1017/CBO9780511618024

**Risk Communication (Dominic Balog-Way, Katherine McComas)**

To earn the social license to operate, it is important to engage in multi-modal risk communication, in which the discourse does not solely aim to align lay opinions with the conclusions of scientific experts; risk judgements are not limited to technical risk assessments; social, cultural and psychological factors are recognized; and the various stakeholders engage through dialogue and deliberation. Risk communication is an iterative process among scientists and non-scientists about risk assessment, risk characterization, risk management and risk policy, which includes purposeful and unintentional messages about risk, and encompasses verbal and non-verbal cues. Responses to strategic risk communication are influenced by the characteristics of the audience, message, source and channel; the social, cultural, economic and political context; and the nature of the risk. Research supports early "upstream" engagement, which allows one to define a baseline of public concerns, values, and perceptions of risk early; build trust early by demonstrating competence, fairness and openness; help messengers to stay ahead of problems; and steer research and development based on knowledge gained early, before decisions are locked in. Research on risk communication related to conventional geothermal systems started in the 80's, but studies on deep geothermal technologies are scarce, and case specific. Scholars at Cornell University have recently collaborated with universities in Switzerland and the U.K. to better understand the features of perception of risk associated with EGS[14].

The first research question that they addressed aimed to identify the positive and negative associations that people make with deep geothermal energy when first learning about the technology. The premise is that low familiarity with a subject shifts the perception of that subject towards pre-existing associations, for example through a representativeness heuristic. Some potential consequences of risk associations include spillover effects (e.g., associating EGS with fracking) and social amplification of risk ripples (e.g., associating nuclear power plants with nuclear weapons). It was found that, despite a low familiarity with geothermal energy, the population sampled in the United States and in Switzerland (2,076 survey respondents) was largely in favor of deep geothermal technologies. Figure 2 shows the average affect scores of the interviewed population on various associations. Fracking and earthquakes were perceived as the most negative associations with EGS, while deep geothermal energy was positively associated with energy independence and sustainable and renewable energy sources. More respondents associated EGS with renewable energies like solar and wind compared to extractive industries like oil and gas. Interestingly, Swiss respondents considered EGS as a renewable energy technology significantly more than US respondents.

---

[13] Barich, A., Stokłosa, A. W., Hildebrand, J., Elíasson, O., Medgyes, T., Quinonez, G., ... & Fernandez, I. (2022). *Social License to Operate in Geothermal Energy. Energies, 15 (1), 139*, doi: 10.3390/en15010139
[14] Cousse, J., McComas, K., Lambert, C., Balog-Way, D., & Trutnevyte, E. (2024). How beliefs about tampering with nature influence support for enhanced geothermal systems: A cross-national study. *Risk Analysis*.

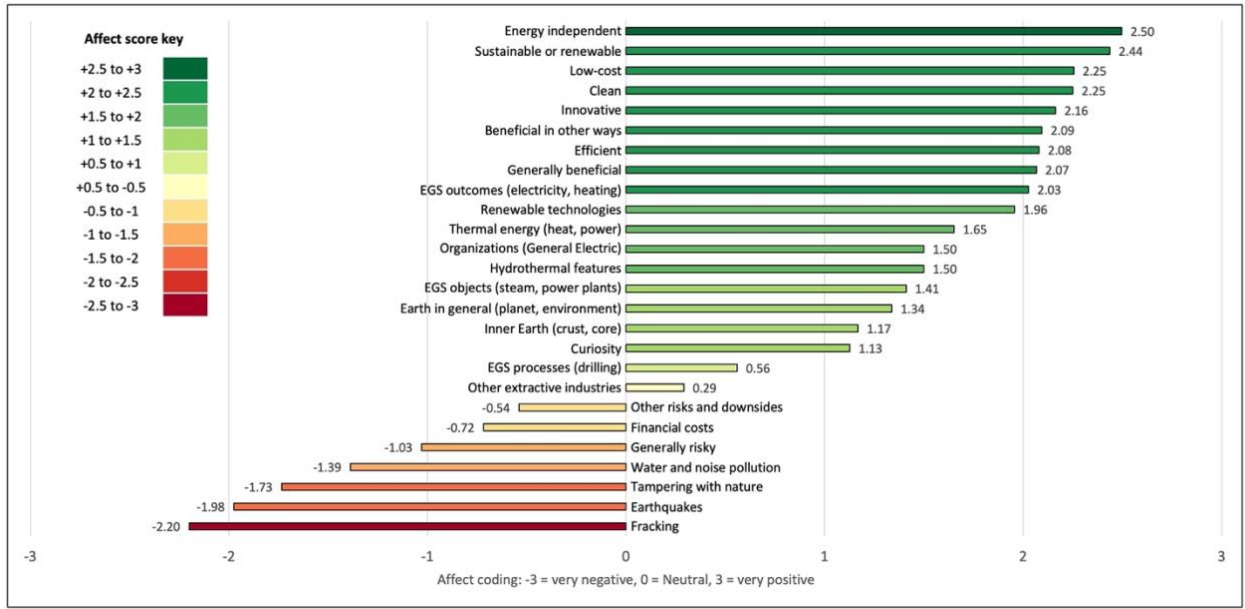

**Figure 2.** Average affect scores assigned by a sample of 2,076 people in Switzerland and in the US regarding various associations made with EGS.

US respondents were significantly more likely than Swiss respondents to say that thinking about EGS made them think about extractive industries, which led to the second research question, which aimed to understand whether prior experience with extractive industries relate to EGS support. It was found that, in general, US respondents were significantly more likely than Swiss respondents to report having had experiences with drilling for oil, mining for coal or drilling water wells. In comparison, Swiss respondents were significantly more likely than US respondents to report having had experiences with drilling for tunnels. No significant differences were noted in relation to experiences with drilling for geothermal among the two countries. In both the US and Swiss surveyed populations, the more experience with drilling industries, the greater the trust with industry, federal agencies, and state agencies. Most respondents indicated that the connection of drilling companies to EGS development was a concern for them. However, this concern was not significantly associated with EGS support or lack thereof.

The third research question focused on the extent to which conceptions of the underground and beliefs about tampering with nature influence support for EGS development. "Tampering with nature" is the "belief that humans influence nature in a negative manner" (Hoogendoorn et al., 2021)[15]. Resistance to technologies may arise from discomfort with tampering with nature. Surveys among the US and Swiss populations indicated that respondents are "in the middle" about EGS as "tampering with nature." EGS is not seen as "tampering with nature" as much as pesticides, genetically modified organisms, or induced seismicity. Responses also suggested curiosity towards underground spaces, which were less viewed as sacred places than pathways for discovery or frontiers for exploration.

Fore and foremost, context influences risk perception and acceptance. For instance, collective visions of place and energy systems differ across spatial scales, with different scales often in conflict. Past studies highlight place meanings and "place-technology fit." Against this backdrop, the fourth research question was to understand how visions of place and energy shape EGS responses. It was found that campus/regional/national focus on innovation and energy leadership through testing and validating a new technology ("data mining") is in competition with the local vision of sustainability leadership, urgent climate action using proven, mature technologies, and opposition to fracking. EGS supporters expressed pride in their county as a leader in sustainability, innovation, and equity. EGS skeptics thought that urgent climate action was incompatible with the practical mix of mature technologies used for EGS development, perceived similarity with fracking, and manifested dread associated with the underground. The underground is thus an important contextual factor influencing how people understand and perceive technologies.

---

[15] G. Hoogendoorn, B. Sütterlin, M. Siegrist (2021). Tampering with nature: A systematic review. *Risk Analysis*, *41*(1), 141-156.

**Economic value or reducing greenhouse gas emissions (Sheila Olmstead)**

Future greenhouse gas (GHG) emissions and local air pollution will depend on choices made by households, firms, universities, and governments. Recent studies show that, the weaker the GHG emission mitigation, the more economic damages from climate change. Therefore, to an economist, the optimal air pollution mitigation plan is the one that is the most efficient in maximizing the net benefits of GHG emission reduction. Economic mitigation is designed to avoid marginal damages: the targeted marginal abatement costs (e.g., for $CO_2$ of $CH_4$) are set equal to the expected marginal benefits in the optimization problem. Multiple approaches integrating models of the global economic and climate systems are used in the current literature to estimate the marginal benefit of GHG mitigation. This so-called the "social cost of carbon" (SCC) began as an academic concept. The United Kingdom was the first national government to propose an estimate for use in policymaking, in 2002. Then, in 2008, a US federal court decision set foundation for the development of US SCC, which is defined as the dollar amount of economic damages per ton of $CO_2$ emissions in a particular year. The SCC is calculated iteratively: after (1) modeling the future socio-economic system of interest, (2) expected GHG emissions from that system are given as input variables to a climate change model that calculates the anticipated temperature increase, precipitation changes, sea-level rise, etc., which allows one to (3) estimate physical impacts of climate changes on humans and ecosystems (the "damage function"), after which, (4) the impacts are monetized. The algorithm is run by increments of 1 additional ton of GHG emissions in the year of interest and calculates the difference in the economic damages for each extra ton of GHG emission, for the whole range of scenarios under study. The output of the model is a distribution of SCC estimates. The SCC rises over time, because damages from pollution typically rise more and more rapidly as pollution increases, leading to greater emission costs on the margin. President Obama created the US Federal Interagency Working Group (IWG) on the Social Cost of Greenhouse Gases in 2008. Figure 3 highlights that, according to IWG SCC simulations, a discount rate of 5% (respectively, 3%, 2.5%) on GHG reduction technologies will <u>in average</u> bring the cost of a ton of emitted $CO_2$ to \$14 (respectively, \$51, \$152). Indeed, results are shown as <u>distributions</u> of the SCC, because different model runs rely on different assumptions about parameters. For example, a model run that assumed rapid emissions growth and a high climate sensitivity would yield greater future damages and thus a higher SCC than a model run with slow growth and lower climate sensitivity. The height of each bar is the fraction of model runs that yielded a particular value for the SCC. For example, using a 3% discount rate, the estimated SCC was around \$20/ton in just over 10% of the model runs. The SCC distributions have a long right-hand "tail," especially at the lower discount rates. Intuitively, there is a small but nonzero probability that damages from $CO_2$ emissions are much higher than the average value. Using a 3% discount rate, for example, the <u>average</u> SCC estimate was \$43, but roughly 10% of the model runs yielded an SCC of \$60 or greater. One way of capturing this "tail effect" is to calculate the 95$^{th}$ percentile value of a distribution: the value that is greater than 95 percent of the model runs for a given discount rate. As the figure shows, the 95$^{th}$ percentile for a 3% discount rate is \$152. For the purposes of regulatory analysis, the interagency working group recommended using the four estimates highlighted in the figure, corresponding to the three average values plus the 95$^{th}$ percentile under a 3% discount rate. These values are now used by agencies across the government in assessing the benefits and costs of policies that are expected to reduce $CO_2$ emissions, including fuel economy standards for cars and trucks, energy efficiency standards for appliances, emission standards for power plants, and so on.

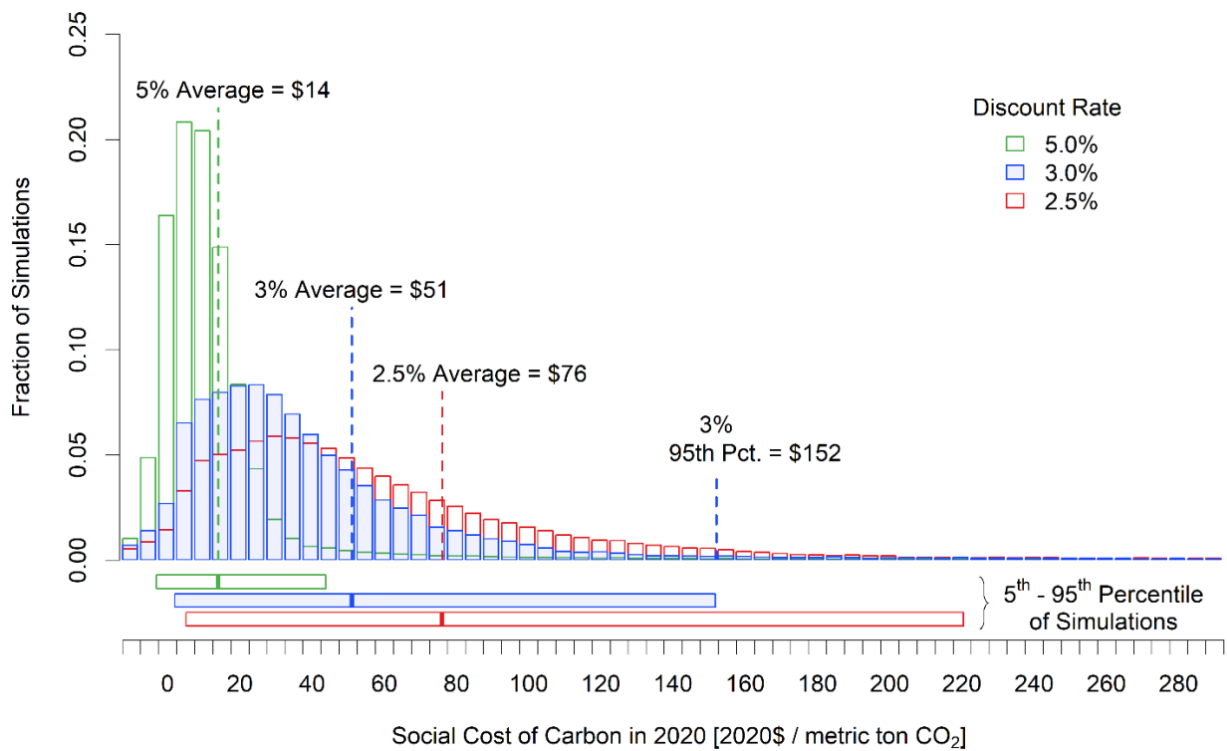

**Figure 3.** U.S. interagency working group Social Carbon Cost simulation results. Source: Interagency Working Group on Social Cost of Carbon, United States Government. *Technical Support Document: Social Cost of Carbon, Methane and Nitrous Oxide interim Estimates Under Executive Order 13990*, February 2021, p. 7.

The IWG developed estimates over the period 2009-2010, for use in federal regulatory impact analysis beginning in 2010. Those estimates were also adopted by Canada, Israel, Mexico and the International Monetary Fund with minimal edits. In March 2017, President Trump rescinded existing federal SCC estimates, then his administration issued its own estimates. The Biden administration temporarily adopted the 2021 IWG numbers and reconvened the IWG to produce a revision, finalized in December 2023. In the latest estimates, the social cost of $CO_2$ is $190/ton for 2020 and $230/ton in 2030 for a discount rate of 2%. This is a lower rate than the rates used in the 2010 analyses, which explains the high estimate SCC. The choice of the 2% central discount rate in the 2023 analysis is linked to the long-term return on government bonds, which dropped in the 90s. The increase in the estimated SCC is also due the increased improved deployment of climate models, and to the use of new damage functions in those models, which reflect new knowledge on climate impacts, such as mortality, agriculture, sea-level rise, and energy consumption. Additionally, probabilistic socioeconomic projections (population, GDP) now inform $CO_2$ emissions projections <u>as well as</u> new stochastic discounting approaches. Indeed, the discount rate declines over time, but stochastically, and varies with economic growth, which is uncertain, because the rate at which we trade current for future consumption depends on how wealthy we think societies will be in the future, which is uncertain. Specifically, the latest probabilistic socioeconomic projections exhibit more internal consistency within the modeling and a more complete accounting of uncertainty, consistent with economic theory (Arrow et al. 2013[16], Cropper et al. 2014[17]) and the National Academies (2017)[18] recommendation to employ a more structural, Ramsey-like approach to discounting that explicitly recognizes the relationship between economic growth and discounting uncertainty.

The economic benefits of reducing GHG emissions are related to the large positive impacts on human health and human capital development, which were convincingly demonstrated in the literature (Zivin and Neidell, 2013)[19]. Indeed, reducing air pollution lessens infant mortality; adult premature mortality as well as respiratory and cardiovascular illnesses; incidence of premature births and low birth weight; and student school absences. It was also shown that reducing air pollution increases student test scores; short-run worker productivity (especially agriculture, construction); and long-run earnings and labor force participation among adults exposed *in utero*, in infancy and in early childhood. According to the Environment Protection Agency (EPA), if earth source heating replaces electricity from natural gas, avoided local and regional air pollution will also generate economic benefits. Fine inhalable particulates that are 2.5 microns or less in diameter ($PM_{2.5}$) account for about 90% of air-pollution-related health damages in the United States. At Cornell, these emissions were already significantly reduced through increases in renewable generation, combined heat and power, and lake source cooling. But significant residual emissions from natural gas electricity generation would be mitigated if direct Earth source heat was used, and those reductions would also have economic value.

**Regulatory challenges: learning from the fossil fuel industry (Sheila Olmstead)**

The legal framework to regulate EGS is still at its infancy, and bridging that gap presents challenges that exhibit similarities with those experienced by local, state, and federal administrations to regulate shale gas extraction: records for regulation, permitting and monitoring are sparse, and novel drilling technologies call for new regulations. Induced seismicity due to drilling, fracturing, and injection is one of the main areas in which federal regulation is needed. Empirical evidence links hydraulic fracturing for oil and gas production with seismicity, mostly from produced water injection. The risk of seismicity induced by hydraulic fracturing is estimated to be higher for EGS than for oil and gas extraction. Prominent examples of EGS induced seismicity include Basel (2006) and Pohang (2017). As a result, risk management and regulation will be key factors to the expansion of EGS in the United States, and ideally, future US regulations will be informed by academic research and legal practices in Europe and elsewhere.

Another ripe area for research is EGS waste management. Fossil fuel operations produce high volumes of chemically complex liquid and solid wastes. Treatment, re-use, disposal, and management have been complicated. For instance, underground injection capacity is limited, especially in the eastern U.S. Disposal via treatment and release from publicly owned treatment works is problematic (Olmstead et al. 2013[20], Shih et al. 2015[21]). Recycling and re-injection can be cost-effective, but volumes have exceeded demand for re-use over time. Naturally Occurring Radioactive Materials (NORMs) in solid waste must be handled by (scarce) selected landfills, which is costly. For example, constituents like chloride and bromide are expensive to remove. Geofluids and waste residual from EGS may have some related challenges, though on a smaller scale in terms of both volumes and constituent concentrations. Median concentrations may be lower in EGS geofluids, but often with higher variability, which could be equally challenging to manage. Heavy metals might be a risk in terms of toxicity with release to the environment (intended or unintended). According to the most current geological surveys, NORMs concentrations in deep geothermal reservoirs are within the range of expected values for rock at that depth. But concerns have arisen because NORMs in Marcellus wastewaters were higher than anticipated. Given the experience with hydraulic fracturing for oil and gas production, it seems opportune to update the regulatory framework for EGS based on scientific research, especially <u>before</u> the technology is deployed at <u>scale</u>, to avoid repeating the regulatory catchup and environmental consequences of the shale boom.

---

[16] Arrow, K. J., Dasgupta, P., Goulder, L. H., Mumford, K. J., & Oleson, K. (2013). Sustainability and the measurement of wealth: further reflections. *Environment and Development Economics*, *18*(4), 504-516.
[17] Cropper, M. L., Freeman, M. C., Groom, B., & Pizer, W. A. (2014). Declining discount rates. *American Economic Review*, *104*(5), 538-543.
[18] National Academies of Sciences, Division of Behavioral, Social Sciences, Board on Environmental Change, & Committee on Assessing Approaches to Updating the Social Cost of Carbon. (2017). *Valuing climate damages: updating estimation of the social cost of carbon dioxide*. National Academies Press.
[19] Graff Zivin, J., & Neidell, M. (2013). Environment, health, and human capital. *Journal of economic literature*, *51*(3), 689-730.
[20] Olmstead, S. M., Muehlenbachs, L. A., Shih, J. S., Chu, Z., & Krupnick, A. J. (2013). Shale gas development impacts on surface water quality in Pennsylvania. *Proceedings of the National Academy of Sciences*, *110*(13), 4962-4967.
[21] Shih, J. S., Saiers, J. E., Anisfeld, S. C., Chu, Z., Muehlenbachs, L. A., & Olmstead, S. M. (2015). Characterization and analysis of liquid waste from Marcellus Shale gas development. *Environmental science & technology*, *49*(16), 9557-9565.

EGS is often perceived as a technology that presents groundwater risks, perhaps because of its association to hydraulic fracturing for fossil fuel extraction, which presents risks of methane, fracking fluid and produced water migration from well casing failures, surface spills, and inadequate produced water disposal. Notably, unconventional oil and gas wells were excluded from the Safe Drinking Water Act's definition of "underground injection" in 2005 National Energy Policy Act, which created controversies. The empirical evidence on the risk of water contamination by fracking for oil and gas production is mixed, with individual case studies that demonstrate contamination, but no large-scale study that suggest systematic contamination. Well integrity is one of the primary known causes of water contamination by hydraulic fracturing. Fortunately, EGS typically comprise less wells than shale gas extraction facilities, which reduces groundwater risks. Furthermore, the vertical separation between the injected geological layers and the aquifer is large in EGS, which also lowers risks of drinkable water contamination.

Fossil fuel extraction facilities require land clearing, well pad creation, and drilling, which amount to significant land disturbance. EGS rely on less wells than shale gas exploitations, which limits land disturbance. Empirical evidence suggests that in Pennsylvania's Marcellus Shale region, the concentration in total suspended solids (TSS) increased in rivers and streams downstream of well pads and wells. It may be possible to address this pollution risk via existing and/or modified stormwater control regulation. The risk is likely lower in an EGS, which requires wastewater disposal only during the creation of flow paths between the injection and production wells, and not during the operation of the facility, during which the heat carrier extracted is reused for injection.

Traffic regulation may have to be updated for EGS, if produced water is to be evacuated for treatment. Produced water transportation has indeed generated heavy industrial traffic at the vicinity of shale gas production sites. Empirical evidence suggests that in the regions of Pennsylvania where Marcellus shale is exploited, truck accidents have increased, and car-only accidents have increased even more, leading to a rise in car insurance premiums. In an EGS, hydraulic fracturing is only needed during construction and not during operation, which should limit the need to truck produced water.

**EGS potential economic and environmental impacts (M. Gillenwater, Tony Ingraffea, Jeff Tester)**

Regulation relies on environmental accounting. Attributional accounting, also called allocational accounting, consists in calculating the 'carbon budgets' of a given entity such as a country, a city or company, which can usually be defined within spatial limits. According to the Greenhouse Gas Management Institute, "Attributional methods generally provide clear rules for identifying a specific set of sources and sinks and allocating 'ownership' or 'responsibility' to different entities" (Brander, 2021)[22]. Allocational accounting is easy to implement for regulatory compliance but does not provide any information to analyze the impacts of specific decisions, which is the scope of consequential methods. Both attributional and consequential accounting are important to greenhouse gas emission mitigation.

---

[22] M. Brander, 2021. The most important GHG accounting concept you may not have heard of: the attributional-consequential distinction, Greenhouse Gas Management Institute, Discussion paper 2021.1

The carbon budget of the United States for energy consumption is mainly attributed to the combustion of fossil fuel such as methane and propane for heating purposes – more so than for electricity production. EGS for heat direct use is one of the technologies envisioned to reduce the US carbon footprint. However, for EGS to contribute to the reduction of GHG emissions, one must consider the entire life cycle of the installation – not only the nature of the fuel being extracted. In an EGS, one of the potential sources of carbon pollution is methane leaks from the wells during drilling, stimulation, and downstream operations. Methane leakage is a known issue from unconventional and conventional gas reservoirs. The combustion of n molecules of methane produces n molecules of carbon dioxide. However, the production and downstream operations of those n molecules of methane release m molecules of methane into the atmosphere unburned as accidental and purposeful emissions. Given the global warming potential of methane of about 80 over a 20-year period, an m = 1.2 about doubles its carbon dioxide equivalency: the effect on global warming is the same as burning the methane twice. In EGS, the risk of methane leakage is site specific, and several mitigation techniques are under development, including high-performance cements for geothermal well integrity (e.g., Bergen et al., 2022)[23] and bacteria-driven methane precipitation.

The path to a more sustainable energy mix seems to be the strategic combination of several alternative technologies to fossil fuels. Since the scalability of EGS is limited, wind, solar and nuclear energy ought to be considered for a successful energy transition. In Iceland, where high-temperature geothermal facilities produce electricity from steam, hot water is used for direct heat, such that geothermal energy is used for dual purposes. It would not be possible to do so in New York State, where temperatures do not exceed 80°C at a depth of 10,000 feet. Dispatching geothermal energy is not efficient, so, it is opportune to develop it for heat direct use as opposed to electricity production, but actual costs of production are still unknown.

## EGS scientific and technological research areas

**Lessons learnt from Utah FORGE (Koenraad Beckers)**

The DOE funded Frontier Observatory for Research in Geothermal Energy (FORGE) was launched in 2014. Initially, five sites were under investigation: West Flank of Coso, CA; Snake River Plain, MD; Fallon, NV; Newberry Volcano, OR; Milford, UT. The Utah FORGE was down selected in 2019 and run by the University of Utah. Over $200M were spent to date on FORGE by DOE Geothermal Technologies Office (GTO). Utah FORGE was recently extended through 2028 as authorized by the Energy Act of 2020 ($80M). The geology of the site is a kilometer-thick alluvium basin fill overlying impermeable crystalline basement rocks. Seven wells were drilled at Utah FORGE, for injection, production and monitoring (see Figure 4). Reservoir temperatures reached around 200°C at 2.5km depth. Significant improvements in Rate Of Penetration (ROP) were achieved well after well, with a ROP in hard and abrasive rock reaching 70ft/h in 2022, which is 5 times the ROP reached at the same site in 2017. For reference, the standard oil and gas ROP ranges between 120 and 200 ft/h. The Efficient Drilling for Geothermal Energy (EDGE) project with Texas A&M[24] was instrumental to enhance drilling techniques at FORGE, through aggressive monitoring, Polycrystalline Diamond Compact (PDC) drill bits improvements, and borehole rugosity mitigation (Figure

---

[23] Bergen, S. L., Zemberekci, L., & Nair, S. D., 2022. A review of conventional and alternative cementitious materials for geothermal wells. *Renewable and Sustainable Energy Reviews*, *161*, 112347, https://doi.org/10.1016/j.rser.2022.112347
[24] Dupriest, F., & Noynaert, S. (2022, March). Drilling practices and workflows for geothermal operations. In *SPE/IADC Drilling Conference and Exhibition* (p. D021S015R001). SPE.

5). As a result, three fiberoptic cables were successfully installed in production well 16B(78)-32. Other, less successful drilling technologies tested at FORGE included particle drilling, insulated pipe drilling and hammer drilling. The 16A(78)-32 injection well was stimulated in three stages: (1) 200ft of open-hole injected with 50 bpm slickwater; (2) 20 ft of perforated cased hole injected with 35 bpm slickwater; and (3) 20 ft of perforated cased hole injected with 35 bpm viscosified fluid. The 16B(78)-32 production well was drilled 300 ft above the 16A(78)-32 injection well to increase heat gains by advection. Although the production well was drilled through the seismic cloud, the connectivity achieved between the two wells was poor, and, in April 2022, the production rate was only 0.5L/s for an injection rate of 13L/s. Further stimulation was conducted April 2024, with a 10-stage plug & perf stimulation in the 16A injection well, and a 5-stage plug & perf stimulation in the 16B production well. The 2024 stimulation stages in the 16A injection well were designed to produce fractures that would intersect with prior 16-A fractures. During the 2024 stimulation operations, several proppants and fluid treatments were tested, the cluster spacing and the number of clusters per stage were varied, and fiber optic measurements were taken along the wells to monitor longitudinal strains, locate fractures induced by stimulation, and identify potential flow paths. The best fracture connectivity was achieved after stimulating both the injection and production wells with proppants. During the 9-hour circulation test of April 2024, a production rate of 21L/s was achieved for an injection rate of 34L/s (62% recovery). This success was confirmed by the 30-day stimulation test with an injection at 27 L/s (10 BPM) in August 2024, which yielded a production rate up to 24 L/s (90% recovery), with a production temperature up to 188°C. Spinner tests in wells 16A and 16B indicated inflow/outflow across most stages.

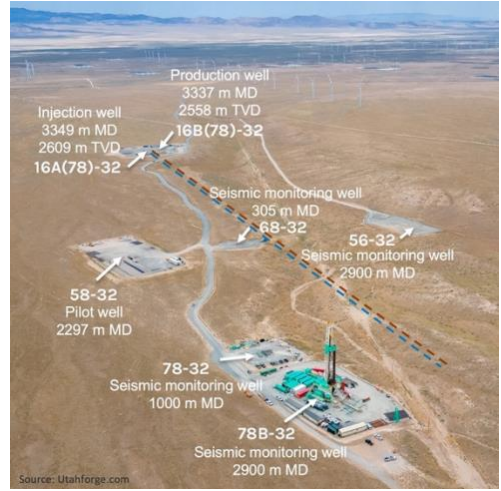

**Figure 4**. The seven wells at the Utah FORGE site.

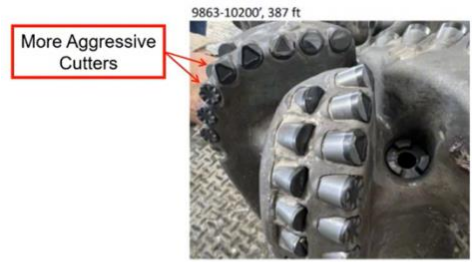

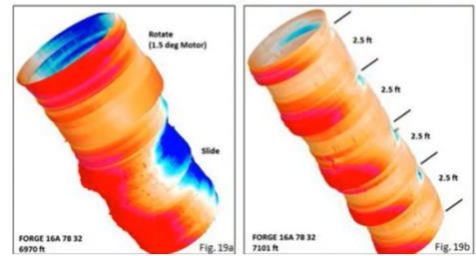

**Figure 5**. Drilling improvements through the EDGE project: PDC bit optimization (top) and well rugosity mitigation (bottom). After Dupriest and Noynaert (2022)[24].

The Fervo Project Cape, directly adjacent to FORGE site, targets a 400 MW$_e$ plant at Project Cape Station. The installation consists of 1-mile-long horizontal laterals at 8,500 ft (2.6 km) depth with a reservoir temperature around 200°C. In July 2024, a 30-day circulation test was conducted, during which two wells were injected at pressures ranging from pressure at 2,000 to 2,300 psi, and fluid was extracted from a third well. The production pressure was 300 to 350 psi, and the required pumping power was about 1.5 MW$_e$. During the test, the production temperature gradually increased to 383°F (195°C). The estimated electric power output was 9.5 MW$_e$ when flowing at 93 kg/s. The project aims to produce 90 MW$_e$ by 2026, and 400 MW$_e$ by 2028. Key results from Fervo's and FORGE long-term circulating tests are shown in Table 1.

**Table 1.** Key results from Fervo's and FORGE long-term circulating tests

|  | Fervo Project Red | Fervo Project Cape | Utah FORGE |
|---|---|---|---|
| Vertical Depth | 7500 ft | 8500 ft | 8000 ft |
| Test Period | April 2023 | July 2024 | August 2024 |
| Test Duration | 37 days | 30 days | 30 days |
| Stimulated Lateral Length | 3000 ft | 5000 ft | 1500 ft |
| Lateral Spacing | 360 ft | 450 ft | 300 ft |
| Injection Rate | 44 L/s | ? | 27 L/s |
| Production Rate | 38 L/s | 93 L/s | 24 L/s |
| Recovery | 85% | ? | 90% |
| Production Temperature | 169°C | 195°C | 188°C |

At Utah FORGE, drilling enhancements, including PDC bit optimization and crew training allowed achieving ROPs of 70+ft/h in hard rock. Lateral well connectivity was the highest when both the injection and production wells were stimulated with proppants. Fiber optic cable deployment was instrumental for monitoring the stimulation treatment. A 24 L/s production rate was obtained from 1,500 ft long stimulated laterals. However, it remains challenging to stimulate tight rock sections. Additionally, fiber optic cables are prone to failure during stimulation, and geophones have a short lifetime at high temperatures. Utah FORGE was extended by four years. During that period, wells 16A and 16B will be subjected to other rounds of stimulation, and additional injection and production wells will be drilled.

**Contemporary geothermal projects other than CUBO and FORGE (Patrick Fulton, Seth Saltiel)**

Geothermal systems installed recently span a large range of depths and resort to various technologies. For instance, Princeton University geo-exchange project involves 2,000 boreholes installed 600 – 850 feet (200-300 m) below ground. These shallow boreholes ought to be combined with ground-source heat pumps to provide district heating and cooling through a closed loop system. The installation comprises over 13 miles (21 km) of distribution piping. Duke University aquifer thermal energy storage system also relies on a low temperature reservoir but consists in storing hot water in permeable sandstone layers at shallow depth. Exploratory drilling has been conducted recently down to 200 meters below ground. The goal is to use this installation for district heating. Delft Subsurface Urban Energy Lab (DSUEL) explores the use of geothermal well pairs for heat direct use. Wells are 3 km deep, where the reservoir temperature is about 80°C. The rocks are softer than in Utah or in the Northeast of the United States, and the reservoir provides shallow heat storage in permeable sandstone layers. A team of researchers at West Virginia University drilled a deep exploratory hole near the university campus in 2023. The project builds upon shallower gas well research conducted at the same site. The EGS envisioned in West Virginia is similar to Cornell's envisioned deep geothermal installation for direct heat production. Public information on the current status of the West Virginia University geothermal project is limited. The ST1 Deep Heat project in Espoo, Finland was led by Aalto University. This EGS aimed to provide district heating from 6.2 – 6.4 km deep wells reaching a reservoir temperature of 120°C. The reservoir rock is impermeable granite, and so, several attempts were made to stimulate the reservoir. Over 49 days in 2018, 18,160 m³ of fresh water without additives or proppants was injected in five stages, each 100–200 m long, at depths of 5.8–6.4 km (vertical depth 5.7–6.1 km) at constant flow rates between 400 – 800 l/min (Kukkonen et al., 2023) [25]. Peak wellhead pressures were ~70 – 90 MPa. Subsequently, in 2020, approximately 7,000 m³ of fresh water was injected into a single open-hole section of a separate well at depths of 4.9–6.2 km (vertical depth 4.8–5.8 km) for 0.5 – 2 hr at flow rates of 200 – 400 l/min attaining wellhead pressures of ~70 MPa. These treatments successfully stimulated natural fractures but did not result in long-term improvement of hydraulic conductivity; therefore, continuation of the project has been put on hold (Kukkonen et al., 2022) [26].

---

[25] Kukkonen, I.T., Heikkinen, P.J., Malin, P.E., Renner, J., Dresen, G., Karjalainen, A., Rytkönen, J. and Solantie, J., 2023. Hydraulic conductivity of the crystalline crust: Insights from hydraulic stimulation and induced seismicity of an enhanced geothermal system pilot reservoir at 6 km depth, Espoo, southern Finland. *Geothermics*, 112, p.102743.

Collab EGS experiments of dense characterization and monitoring focuses on meso-scale testing, utilizing the access to tunnel walls at relevant stress conditions. Meso-scale EGS Experiments at SURF (Homestake Mine, South Dakota) targeted planar hydraulic fractures in the first experiment location, then shear stimulation of existing fractures in the second experiment at a shallower test bed which was later hydraulically fractured after shear propping proved unsuccessful (Kneafsey et al., 2024)[27]. Hydraulic fracturing experiments were accompanied by thorough monitoring and jet observations (Fu et al., 2021)[28]. Distributed temperature sensing (DTS) allowed collection of temperature time series with injection. Three-dimensional strains were measured across fractures to monitor slip during injection. Micro-earthquake observations revealed that less than a quarter of slip is co-seismic. Fracture planes were identified through micro-seismic monitoring and water jetting observations. Discrete Fracture Networks were mapped from core image analysis and micro-earthquake location. Ambient noise stacks recorded in the borehole Distributed Acoustic Sensing (DAS) system was used to identify fracture locations (Li et al., 2024)[29]. Shear stimulation was monitored through multi-modal sensing fibers. Self-propping by shear dilation or gouge produced by frictional wear proved unsuccessful. Electrical resistivity tomography (ERT) images highlighted stress effects as compressed cracks closed: resistivity increased in surrounding rock with increasing pore pressure in the fracture, and conductivity increased in the fracture as hydraulic propping was achieved (Johnson et al., 2024)[30]. Collab efforts allowed assessing EGS conditions spatially and temporally through monitoring micro-seismicity and acoustic emissions, temperature transients, observations of fluid jetting, injection pressure and flow rate, estimates of surface and fault strains, and characterization of lithology and pre-existing faults and fractures (from core and borehole images). The instrumentation system allowed characterization of the fracturing and seismicity that took place during the tests (hydraulic, shear, thermal), interpretation of aseismic, pre-, co, and after seismic slips, estimation of rock strength, measurement of thermal and hydraulic conductivity, and assessment of the 3D structure of the reservoir (seismic velocity, fracture density and orientation). It was found that stress and temperature gradients affected fracture trajectories, and that the use of external proppants was necessary for fractures to remain hydraulically open. Indeed, shear displacement was observed but did not enhance fracture permeability, and existing healed and mineral filled fractures did not open or were pressurized below the minimum principal stress. Note however that naturally conductive fractures were avoided over excess flow concerns. Results also showed that flow paths were distributed and evolved dynamically, and that flow was observed in unexpected locations, limiting production. Despite constant injection rates, 'production' rates fluctuated at many locations. Lastly, it was found that micro-earthquakes did not necessarily correlate with flow paths. The meso-scale experiments provide a highly detailed picture of EGS stimulation processes that is much more limited with standard field-scale geophysical monitoring techniques.

---

[26] Kukkonen, I.T., Heikkinen, P.J., Sinisaari, M., Rytkönen, J., Karjalainen, A., Malin, P., Giese, R. and Kueck, J. (2022), St1 Deep Heat Project: Geothermal energy from 5-6 km in the continental crust. *European Geothermal Congress 2022.*

[27] Kneafsey, T., Johnson, T., Burghardt, J., Schwering, P., Frash, L., Roggenthen, B., Hopp., ... & EGS Collab Team. (2024). The EGS Collab Project – Summaries of Experiments 2 and 3: Experiments at 1.25 km depth at the Sanford Underground Research Facility. https://escholarship.org/uc/item/43k0p074

[28] Fu, P., Schoenball, M., Ajo-Franklin, J. B., Chai, C., Maceira, M., Morris, J. P., ... & EGS Collab Team. (2021). Close observation of hydraulic fracturing at EGS Collab Experiment 1: Fracture trajectory, microseismic interpretations, and the role of natural fractures. *Journal of Geophysical Research: Solid Earth*, *126*(7), e2020JB020840.

[29] Li, D., Huang, L., Zheng, Y., Li, Y., Schoenball, M., Rodriguez-Tribaldos, V., ... & Robertson, M. (2024). Detecting fractures and monitoring hydraulic fracturing processes at the first EGS Collab testbed using borehole DAS ambient noise. *Geophysics*, *89*(2), D131-D138.

[30] Johnson, T. C., Burghardt, J., Hammond, G. E., Karra, S., Jaysaval, P., Rosso, K., & Hyman, J. D. (2024, June). Electrical Resistivity Tomography based monitoring of stress perturbations to optimize placement of high-precision strain meters. In *ARMA US Rock Mechanics/Geomechanics Symposium* (p. D032S040R011). ARMA.

SuperHot Rock (SHR) rock reservoirs are at temperatures that exceed 375°C. To alleviate the cost of SHR geothermal energy, it was recently proposed to use ultra-high enthalpy heat carriers, which could enable higher capacity facilities, reduce the number of wells per MW produced, and increase turbine efficiency. Pure water reaches supercritical conditions at 375°C and pore pressures of 22 MPa (Figure 6). Native supercritical fluids in geothermal wells are rare, so, SHR geothermal systems must be engineered to allow circulation of supercritical fluids. Initial lab-scale experiments of hydraulic fractures induced by supercritical fluids in semi-ductile rocks such as granite or basalt form clouds of fractures, in comparison to localized fractures that are typically created by classic hydraulic fracturing operations in brittle rock. Around twenty wells have been drilled world-wide in SHR conditions, including two wells in Iceland (Krafla and Reykjanes), two wells at The Geysers in California, four wells in Larderello in Italy, and one well in Kakkonda in Japan. These experiences offer insight into the ability to drill into and extract heat from these reservoir conditions. The 1995 - Kakkonda WD-1a well in Japan was an early, intentional, conventionally drilled SHR well, reaching a reservoir at 500 °C. The 2010 – NW Geysers EGS demonstration in California used an open hole stimulation at 25 kg/s at 400 °C, which triggered up to 42 micro-seismic events per day. The 2009 - Iceland IDDP-1 well intersected rhyolite magma unexpectedly at 2104 m. A 16-month 30 MW flow test was conducted. Native permeability was found in parts of the reservoir that were above 500°C. A core extracted from the 2017 - Iceland IDDP-2 well reached 410°C and a well inclined at 30° traversed rock above 400°C. A temperature of 520°C was reached in the low permeability reservoir of the 2018 - Italy – DESCRAMBLE Vennelle-2 well. No stimulation has been attempted there yet, but follow-up Krafla Magma Testbed with deep injection is currently in planning. Projects in New Zealand, with a goal of commercial generation in the near future, and Newberry volcano in Oregon, are also underway.

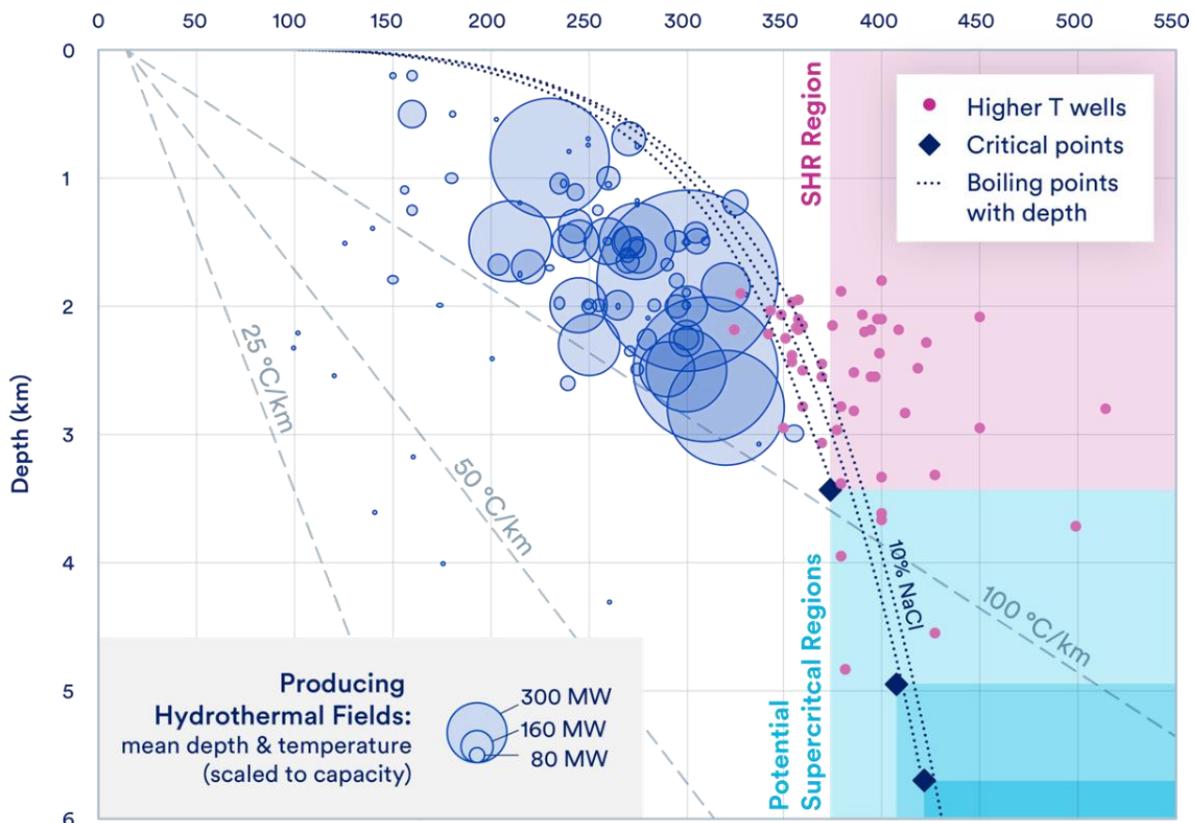

**Figure 6.** Producing geothermal fields and SuperHot Rock wells. After Cladouhos and Callahan (2023)[31].

---

[31] Cladouhos, T. T., & Callahan, O. A. (2023). Heat extraction from superhot rock: a survey of methods, challenges, and pathways forward. *Transactions—Geothermal Resources Council*, 2804-51.

**Recent advances in computational geomechanics for EGS (Chloé Arson)**

Field data interpretation, stimulation planning, and EGS design require solving inverse problems, simulating fracture propagation and fracture flow in heterogeneous media, and estimating thermal drawdown, pore pressure distributions, surface displacements and potential fault slip. It is important to use the numerical method that is the most appropriate to the scale and variables of the problem at hand, and to the accuracy targeted for the solution. In the past five years, computational advances for EGS have focused on hydraulic fracturing simulation, poromechanical pressurization modeling, and geomechanical optimization and forecasting assisted by Artificial Intelligence (AI).

*Hydraulic fracturing simulation through the dynamic insertion of discontinuities.* According to Griffith's theory, fracture propagation is governed by the mechanical work needed to overcome the cohesion of material faces induced by molecular attraction forces. Therefore, according to Griffith's theory, when an elementary segment of fracture propagates in a solid, the total potential energy of that material changes by an elementary decrease of elastic deformation energy and an elementary increase of surface energy. Fracture propagation thus occurs if the elementary change of elastic deformation energy exceeds the energy released per surface area of elementary segment of fracture. To determine whether or not a fracture propagates, one must know the expression of the elastic deformation energy of the non-fractured material. When the solid of interest is not linear elastic, which is the case when dissipation occurs through plastic deformation or pore fluid flow, the expression of the potential energy of the non-fractured medium is not straightforward. In numerical methods in which fractures are represented by discontinuities (i.e., two surfaces that separate), the fracture propagation criterion is governed by a traction-separation law. When the force applied at the fracture faces is expressed as a function of the relative displacement between fracture faces, the area below the force curve represents the energy released. When that energy released exceeds a critical value that corresponds to the energy needed to separate two fracture faces over an elementary length, then, the fracture propagates. Because traction-separation models reflect the response of a cohesive fracture (by contrast with an adhesive fracture, for example), models governed by traction-separation laws are usually referred to Cohesive Zone Models (CZMs). Intrinsic CZMs are deployed in the models from the onset of the simulation, such that the traction-separation law reflects an elastic response (deformation but no debonding) before fracture propagation, followed by a post-peak response that reflects the residual cohesion of the material as the fracture propagates. In extrinsic CZMs, cracks are inserted in the model only when they start propagating, such that the traction-separation law does not exhibit any elastic regime. CZMs recently proposed to simulate fracture propagation in geomechanical reservoirs were extrinsic and were implemented in the eXtended Finite Element Method (XFEM). In the Finite Element Method (FEM), field variables such as displacements and pore pressures are solved at discrete points in a mesh (usually, the element nodes), and the overall solution is interpolated by means of linearly independent interpolation functions (usually, polynomial or trigonometric functions). In the XFEM, discontinuous interpolation functions allow calculation of displacement jumps and pressure gradient jumps within elements. The governing equations of hydraulic fracturing include momentum balance equations for the pore fluid and for the composite made solid rock and pore fluid, as well as mass balance equations in both the non-fractured continuum and the fractures. To avoid discontinuity of the pore pressure field across the fractures, some researchers proposed to estimate the flow of two non-miscible fluids in a fracture by modeling fractures as embedded discontinuities (cohesive zone elements), and to solve the mechanical problem of fracture propagation by discretizing the displacement field with the XFEM[32]. Another group implemented an algorithm in the XFEM to simulate multi-scale mixed mode fluid-driven fracture propagation in transversely isotropic porous media[33] (Figure 7). In that algorithm, a non-local anisotropic damage model is coupled to a traction-separation law that governs the mechanical behavior of cohesive elements. The transition from continuum damage to cohesive fracture is done by dynamically inserting cohesive segments once the weighted damage exceeds a certain threshold. Another recent study presents an XFEM discretization of thermo-poro-mechanical governing equations to simulate two-phase fluid flow in fractured porous media and thermal fracture propagation in unsaturated porous media[34]. Unfortunately, the XFEM fails to simulate fracture intersection and fracture bifurcation.

---

[32] Ren, G. and Younis, R. M. (2021). "An integrated numerical model for coupled poro-hydro-mechanics and fracture propagation using embedded meshes." *Computer Methods in Applied Mechanics and Engineering*, 376, 113606
[33] Jin, W., & Arson, C. (2020). Fluid-driven transition from damage to fracture in anisotropic porous media: a multi-scale XFEM approach. *Acta Geotechnica*, *15*, 113-144.

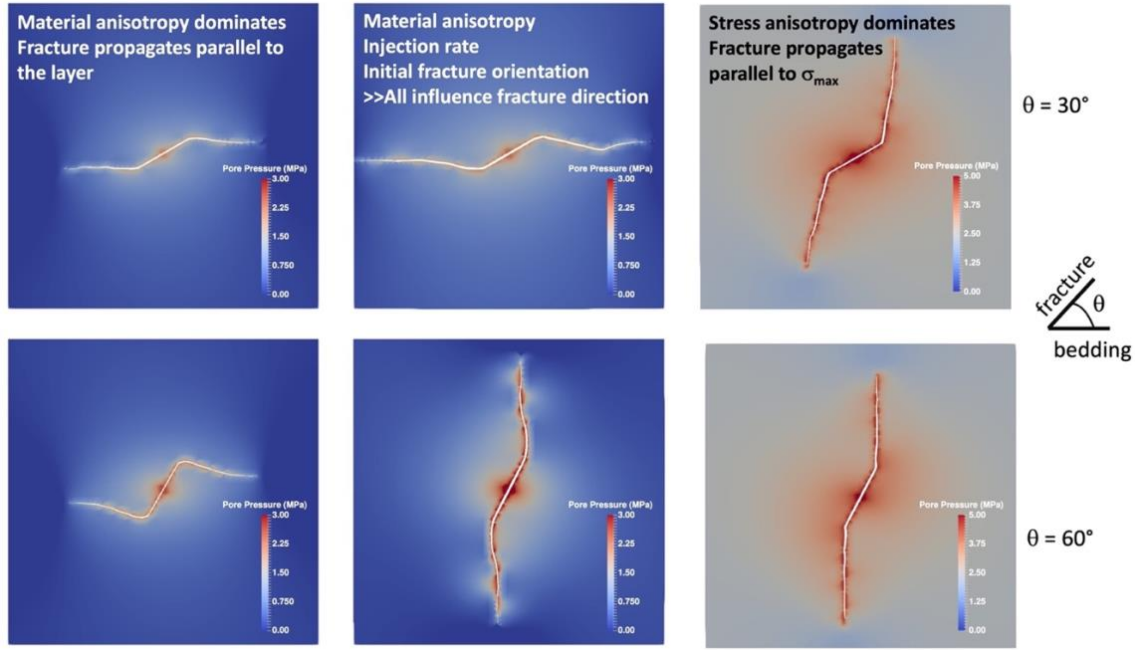

**Figure 7**. Example simulation results obtained with the XFEM approach proposed by Jin and Arson (2020)[33]. A saturated transverse isotropic porous medium is subjected to an injection flow rate in a pre-existing fracture oriented at an angle in reference to the bedding plane. The outer boundaries of the domain are drained. In the first test (left), a zero-displacement condition is applied on the outer boundaries and the injection rate is 10 mm$^2$/s. In the second test (middle), the boundary conditions are the same, and the flow rate is 20 mm$^2$/s. In the third test (right), the flow rate is 20 mm$^2$/s but this time, the outer boundaries are subjected to a biaxial state of stress (vertical compression higher than the horizontal compression). The pore pressure distribution and the fracture path are shown at the end of the simulations (at t = 0.02s).

*Hydraulic fracturing simulation with the Phase Field Method.* Hydraulic fracturing is, in fact, a moving boundary problem: the fracture tip advances in the solid matrix, the displacement jump that separates fracture faces is not uniform, and the fluid that invades the fracture may lag behind the tip, creating a capillary interface in the fracture. The phase field method has recently emerged as an attractive technique to simulate fracture propagation, because it relies on a simplified representation of interfaces: auxiliary variables are introduced to label spatial positions as one phase or another. The model is based on thermodynamic equations that provide the evolution laws of the field variables (such as stress, pore pressure and temperature) and those of the phase variables. The phase field method was initially developed to simulate the flow of immiscible fluids. It is routinely used to model dissolution and precipitation. The variational approach to fracture mechanics was first presented by Bourdin et al. (2008)[35], which led to the extension of the phase field method to fractures, which are seen as a phase change from intact solid to damaged solid. When the phase field equations are discretized in the FEM, each element is assigned one or more phases (for example, intact/damaged porous solid, liquid/gas fluid). In a problem of hydraulic fracturing, each element represents a fluid-saturated porous solid, which represents a highly permeable porous fracture when the damage variable is close to 1, and an intact porous solid when the damage is zero. The fracture is represented as smeared damaged zone, the size of which is dependent on an internal length parameter. Phase field models are computationally efficient and robust because they do not require dynamic remeshing, and they can be formulated to avoid mesh dependency. As an example, the Integrated Phase-Field Advanced Crack Propagation Simulator (IPACS) is an open-source software written in C++ and based on the open-source finite element framework deal.II, which couples geomechanics with flow in porous media in which (possibly multiple) fractures are described using a phase-field technique[36]. Recent progress in phase field methods has allowed simulation of hydraulic fractures as a single-phase propagation problem[37]. Phase-field approaches were used successfully to simulate the path of a hydraulic fracture that interacts with natural fractures represented by a pre-defined damaged zone[38,39,40,41]. Hydraulic fracturing from geothermal lateral wells was successfully simulated with the PFM[42] (Figure 8).

---

[34] Khoei, A. and Mortazavi, S. (2020). "Thermo-hydro-mechanical modeling of fracturing porous media with two-phase fluid flow using x-fem technique." *International Journal for Numerical and Analytical Methods in Geomechanics*, 44(18), 2430–2472.
[35] Bourdin, B., Francfort, G. A., and Marigo, J.-J. (2008). "The variational approach to fracture." *Journal of elasticity*, 91, 5–148.

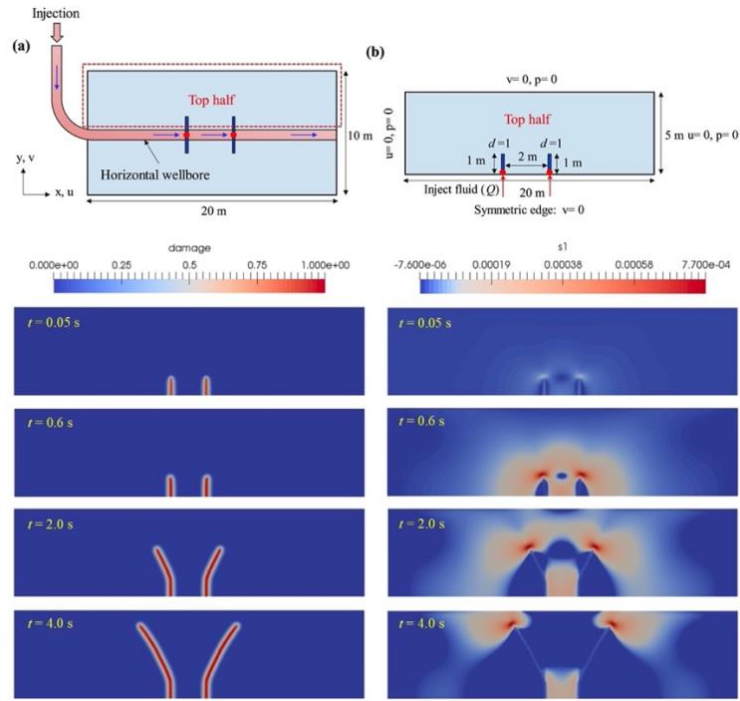

**Figure 8** Simulation of hydraulic fracture stimulation from a lateral well with the PFM, from Li et al. (2021)[42]. The bottom figures show the damage and maximum principal stress evolutions for a flow rate of 3.0 x $10^{-4}$ m$^2$/s and a fluid viscosity of 0.001 Pa.s.

---

[36] Wheeler, M. F., Wick, T., and Lee, S. (2020). "Ipacs: Integrated phase-field advanced crack propagation simulator. an adaptive, parallel, physics-based-discretization phase-field framework for fracture propagation in porous media." *Computer Methods in Applied Mechanics and Engineering*, 367, 113124.
[37] Chukwudozie, C., Bourdin, B., and Yoshioka, K. (2019). "A variational phase-field model for hydraulic fracturing in porous media." *Computer Methods in Applied Mechanics and Engineering*, 674 347, 957–982.
[38] Yi, L. P., Waisman, H., Yang, Z. Z., & Li, X. G. (2020). A consistent phase field model for hydraulic fracture propagation in poroelastic media. *Computer Methods in Applied Mechanics and Engineering*, 372, 113396.
[39] Li, H., Lei, H., Yang, Z., Wu, J., Zhang, X., & Li, S. (2022). A hydro-mechanical-damage fully coupled cohesive phase field model for complicated fracking simulations in poroelastic media. *Comp. Meth. Appl. Mech. & Eng.*, 399, 115451.
[40] Burbulla, S., Formaggia, L., Rohde, C., & Scotti, A. (2023). Modeling fracture propagation in poro-elastic media combining phase-field and discrete fracture models. *Comp. Meth. Appl. Mech. & Eng.*, 403, 115699.
[41] Sarmadi, N., Mousavi Nezhad, M., & Fisher, Q. J. (2024). 2D Phase-Field Modelling of Hydraulic Fracturing Affected by Cemented Natural Fractures Embedded in Saturated Poroelastic Rocks. *Rock Mech. & Rock Eng.*, 1-28.
[42] Li, M., Zhou, F., Yuan, L., Chen, L., Hu, X., Huang, G., & Han, S. (2021). Numerical modeling of multiple fractures competition propagation in the heterogeneous layered formation. *Energy Reports*, 7, 3737-3749.

*Poromechanical pressurization models.* To understand the impact of non-isothermal fluid flow through geothermal reservoirs, the weak form of thermo-poro-mechanical governing equations was transformed into the Laplace space where they were solved by the Galerkin finite element method to simulate changes of pore pressure and temperature at the wall of cavities embedded in fluid-saturated porous media in the unsteady state regime[43]. The advantage of the proposed Laplace-Galerkin method is that it does not require any time step, since it is solved in the Laplace space where time derivative terms vanish. This presents major savings in computational cost. A three-dimensional thermo-poroelastic model of hydraulic fracturing was implemented in a mixed finite volume/finite element method (FVFEM) in which fractures are modeled with the displacement discontinuity method (DDM)[44]. The model was used successfully to model fluid circulation in hot dry rock and make recommendations for EGS design. It was found that except for fluid pressurization and stress relief caused by cooling down, the stress shadowing effect induced by propagating fractures constitutes an important mechanism in triggering slippage/microseismic events, and that for short-term hydraulic stimulation, temperature may have a second-order effect on fracture propagation.

*AI optimization and forecasting.* In the 2010s and 2020s, several data-driven approaches successfully predicted the ROP. Random Forest algorithms proved to be some of the best ROP predictors when using drilling parameters extracted at 7,300 different depths of FORGE Well 58-32[45,46]. Extra tree, gradient boosting regressor and light gradient boosting regressor algorithms also performed well. Taking drilling operational parameters such as Weight On Bit (WOB), pump flow rate (GPM), Revolutions Per Minute (RPM) and rock strength as inputs, Hegde and Gray (2018)[47] optimized coupled data-driven models of ROP, Mechanical Specific Energy (MSE) and Torque On Bit (TOB) with a metaheuristic algorithm. It was found that maximizing the ROP can lead to drilling programs with detrimentally high MSE, while optimizing the MSE has the advantage to yield drilling programs with high ROP and low TOB. Alali et al. (2021)[48] proposed a method to adjust the WOB, GPM and RPM in real time to optimize the ROP. The drilling plan is first established based on historical best ROP at the same site and same depth. During the drilling of a new well, the WOB and RPM optimized based on historical data are subjected to a small perturbation for a 5 ft section, and the ROP is calculated. The combination of the WOB and RPM that provides the highest ROP is then selected to drill the next 5 ft section of the well. Machine Learning was also used to minimize the thermal drawdown of geothermal reservoirs. While classically, thermal drawdown curves are established from analytical solutions of heat transfer problems for single idealized circular fractures[49], deep learning[50] has allowed the optimization of injection temperature, injection rate, extraction well pressure and well distance from a ground truth dataset generated by FEM simulations that provided estimations of cold thermal front in the reservoir rock. Recently, diffusion models that were initially adopted in Natural Language Processing (NLP) have allowed stress estimation under a variety of boundary conditions. A Conditional Generative Adversarial Network (cGAN) was trained to predict the von Mises

---

stress field in a 2D domain of variable geometry, based on three input images[51]: the image of the geometry, which contains different pixel values for parts of the solid that are subjected to fixed displacements, an image that maps the intensity of the horizontal forces applied to the domain, and an image that maps the intensity of the vertical forces applied to the domain. That cGAN was trained for a specific linear elastic material based on a data set with a size of the order of $4 \times 10^4$ data quadruplets. Following a similar strategy, a cGAN was trained based on single inputs to estimate the von Mises stress in a biphase composite material of random microstructure, by augmenting the input image of the geometry with annotations of different colors to encode the boundary conditions[52] (Figure 9). In total, 2,000 microstructure images were used for training and testing, and the cGAN accurately estimates the stress field for a variety of boundary conditions, including a compression followed by unloading. The constitutive properties of each phase were non-linear and non-elastic, but the same throughout all the microstructures. A progressive transformer diffusion model based on a cGAN was trained with Molecular Dynamics simulation results to estimate the von Mises stress in a homogeneous domain with flat (crack-like) cavities[53]. The input to the cGAN is an image that encodes the geometry and boundary conditions of the problem. Cascading U-Nets are used, which allows using a very small dataset of size 1,000 only.

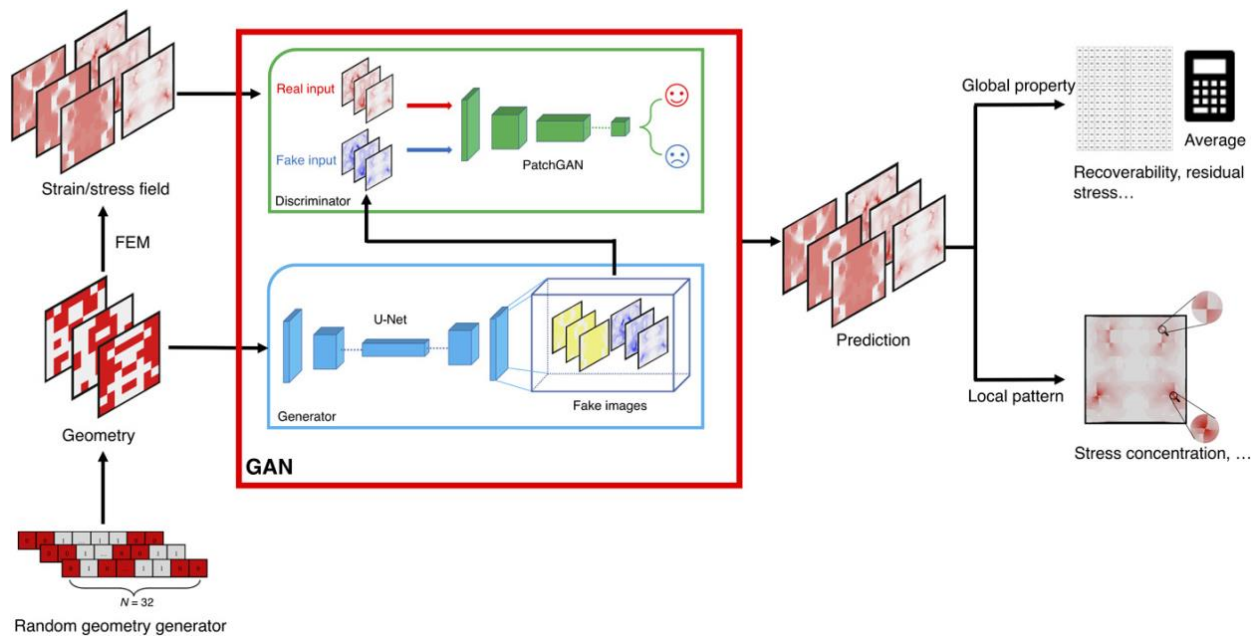

**Figure 9.** cGAN to predict von Mises stress, after Yang et al., (2021)[52]. The input is one annotated image (an image of the geometry that encodes loading and boundary conditions). The size of the data set is 2,000 annotated images.

---

[51] Y. Jadhav, J. Berthel, C. Hu, R. Panat, J. Beuth, and A. B. Farimani. Stressd: 2d stress estimation using denoising diffusion model. Computer Methods in Applied Mechanics and Engineering, 416:116343, 2023.
[52] Z. Yang, C.-H. Yu, and M. J. Buehler. Deep learning model to predict complex stress and strain fields in hierarchical composites. *Science Advances*, 7(15):eabd7416, 2021.
[53] M. J. Buehler. Predicting mechanical fields near cracks using a progressive transformer diffusion model and exploration of generalization capacity. *Journal of Materials Research*, 38(5):1317–1331, 2023.

# EGS case study: Cornell University

**Cornell climate goals (Sarah Carson, Stacey Edwards, Cole Tucker)**

Cornell University uses 2.9 trillion BTU per year to operate about 150 buildings spread over 2,300 acres. Put into perspective, Cornell's Ithaca campus uses 0.1% of New York State's peak power. The synchrotron power consumption currently accounts for 12% of the campus electricity use. Most of the electricity consumed by Cornell's Ithaca campus is self-generated via an efficient natural gas fired combined heat and power plant. About 95% of the steam and hot water supply are byproducts of electrical production, while 98% of the chilled water supply is produced via efficient Lake Source Cooling. In 2007, the President of Cornell University signed a carbon commitment to develop a climate action plan to achieve carbon neutrality. The plan describes actionable items to reduce Cornell's carbon footprint, and emphasized the goal to expand research and education activities to limit and mitigate greenhouse gas emissions. Actions were initially taken voluntarily. External regulatory obligations are now overtaking voluntary commitments, in particular with the New-York State climate law (2019), Ithaca Energy Code Supplement (2021) and mandates related to federal funding via the Inflation Reduction Act and the Department Of Energy (2023). Figure 10 shows Cornell's most recent greenhouse gas emission inventory. The 2023 emissions are 42% lower than the 2008 baseline, mostly because of a shift to combined heat and power which increased the efficiency of generating campus energy by 50%, investment in energy conservation that has offset growth, and because carbon removal technologies were deployed.

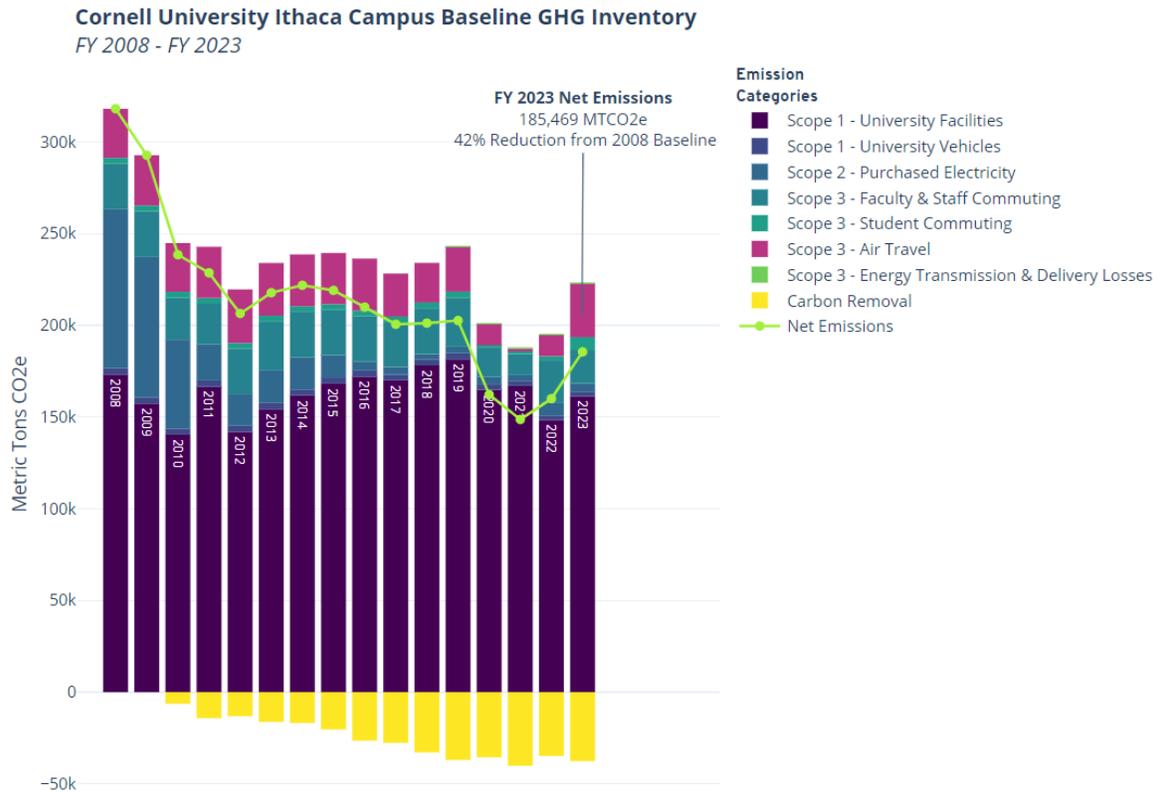

**Figure 10.** Cornell's greenhouse gas emissions inventory between 2008 (after the signature carbon commitment by Cornell University President) and 2023.

Emission reduction key actions at Cornell University include the construction or retrofitting projects for high-efficiency buildings, the replacement of the steam heat distribution system by a hot water heat distribution, the use of carbon free electricity, and the use of carbon free heat. Energy conservation projects such as energy-efficient constructions and retrofits have been undertaken at Cornell since the 90s. In 2024, four rooftop exhaust stacks were installed on the biotechnology building to recover heat. Despite the increase of the campus gross square footage, the Ithaca campus energy use has been stable, which indicates that emission reduction actions have had a positive impact. For instance, the Space Science Controls upgrade divided the energy consumption of the building by two, leading to savings that are expected to yield a return on investment after 6 years. The Biotech Heat Recovery project allows harvesting 50% of the exhausted heat, leading to savings that will yield a payback in 9 years. The replacement of the steam heat distribution system by a hot water heat distribution is expected to cost over $350M, because heat is currently distributed via an expansive and partially aging steam system, made of 12 miles of buried steam and condensate pipes. Additionally, some buildings require extensive, in-space upgrades to accommodate incoming temperatures of 130F. This is because the new heat distribution system targets a 180F supply temperature (temperature at which most facilities operate) with at least a 140F return. The upgrade of the heat distribution system depends on the choice that is made for the renewable heat source: Earth Source Heat (EGS) vs. shallow ground heat pumps. Alternatives for heat are critical because the transformation of the electric grid in New-York State will increase the electricity peak load to 2-3 times current levels, which will significantly lower reliability margins. Recent analyses have shown that, to satisfy the winter heating demand in 2050, Cornell will need to secure the capacity to produce the equivalent of 40GW of carbon-free energy per hour.

The BIG RED (Building an Innovative Grid for Reliability, Efficiency and Decarbonization) energy transition plan addresses the utilities infrastructure portion of the President's Climate Commitment to make the Ithaca campus a net-zero carbon campus by 2035. It excludes buildings outside district energy, campus process gas, university vehicles, commuting, and air travel. Cornell's "fossil fuel free" district energy system will comprise solar power (the equivalent of 20% of the annual load is already produced by solar panels; 80% of the annual load is under contract); electricity from the New York State grid (which is assumed to be zero emissions by 2040); cooling provided by a mix of Lake Source Cooling, chillers, and heat pumps; geothermal heat (shallow geoexchange or deep direct use); hot water distribution (instead of steam); unmet gaps for "process loads" (e.g., humidification, sterilization, steam processing); and a back-up power system in case of grid outages (the source of which is still to be determined). One of the main challenges is to address the peak load of electricity or heating. There are less than 100 hours in the year where heat usage exceeds 80% of the peak. However, such weather conditions may coincide with low availability of renewables (wind/solar). Thermal storage could help, but increased capital costs are expected to meet the peak loads. The same challenges exist for chilled water. Some other key successes in reducing Cornell University carbon footprint have been Lake Source Cooling (which reduced the total campus energy use by 10%), the adoption of LED lighting (which reduced the total campus electrical use by 4%), and the construction of solar farms (which currently produce 30MW).

EGS fits in the BIG RED energy transition plan because it requires low power consumption, minimizes refrigerants, can serve for base heat load but is also dispatchable, and pairs well with Lake Source Cooling. More data is needed to calculate the actual power consumption to operate an EGS (because pumping power varies with well casing diameter and fluid velocity), estimate the actual heat production, assess whether a heat pump will be required, predict the reservoir thermal drawdown, design a system to make up the water loss, and evaluate the accessibility of clean power in New York State. To meet the net-zero carbon goal by 2035, Cornell needs to have demonstrated the EGS technology by 2029.

**Economic value of avoided GHG emissions from Cornell's ESH project (Sheila Olmstead)**

The estimated $CO_2$ emissions avoided by EGS amount to 96,700 tons/year according to the 2016 Climate Neutral Campus Alternatives Analysis. $CO_2$ damages using central estimates for 2020 amount to $51/ton (IWG 2021) or $190/ton (EPA 2023). Avoided yearly damages are thus estimated to be between $4,931,700 (IWG 2021) and $18,373,000 (EPA 2023). Note that these estimates do not include methane (valued at ~$1,600/ton for 2020 emissions) and do not account for any life-cycle emissions associated with earth source heating.

**Cornell University Borehole Observatory (CUBO) (Terry Jordan)**

The Cornell University Borehole Observatory (CUBO) was drilled in 2022. Temperatures measured 15 months after drilling spanned from 76C at 8,680 ft depth to 79C at 9400 ft, allowing extrapolation to >80C (162F-176F) at the greatest depth drilled. These temperatures bracket the sedimentary rock and basement rock geothermal reservoir targets, and are sufficient to produce heat for direct use. Borehole breakout observations indicated that the maximum compressive horizontal stress is oriented North-East / South-West. The magnitudes of the horizontal stress components were determined by modular dynamic tests, while the vertical stress was calculated from the mass densities of the geological layers. From those stress analyses, it is expected that opening fractures will propagate vertically at basement depth, in planes parallel to the North-East / South-West direction.

Thermal Anomalies were repeatedly observed through measures taken in the first week after drilling ended, which indicated the exchange of fluid between the borehole and the solid rock. It is hypothesized that this fluid exchange is due to the presence of natural fractures that constitute advection zones, as shown in Figure 11. Note however that water flow out of the well was low, even when a final experiment attempted to stimulate such a flow. Electrical resistivity at multiple distances from the wall of the borehole indicated that there are multiple zones in which there is effective fluid connectivity from the borehole wall up to about 1 m into the rock mass. In *thin* zones, natural conditions include some layers with effective fluid connectivity *and* the drilling activities induced a small degree of flow of water through the rock. The main question to answer is whether or not adequate flow of water can be induced in the rocks of adequately high temperature to harvest the magnitude of heat needed to achieve the baseline goal of heating large campus buildings. Rocks in which this could be achieved would be potential "geothermal reservoirs."

Potential geothermal reservoirs at CUBO include sedimentary rock in the Galway and Potsdam formations (between 8,700 and 9,400 feet deep, with temperatures between 76C and 79C) and basement rock (below 9,400 feet, with temperatures above 79C). The sedimentary rock is made of quartz, dolomite, and traces of other minerals, and contains abundant horizontal planes of weakness, spaced approximately every meter in the Galway formation and every 0.3 meter in the Potsdam formation. Natural water transmission was noted in thin zones: in two zones in the Galway formation, and in the lower Potsdam formation. However, the porosity of the sedimentary rock is low: 1% in the Galway formation, and 2.6% in the Potsdam formation. These sedimentary rocks have high strength and could be challenging to stimulate by hydraulic fracturing. The properties of the basement rock were found to be highly variable within the 390 ft (120 m) drilled into the basement. The mineralogy of the basement rock is complex: about 40% phyllosilicates, 10-38% quartz, 20-55% K-feldspar. The basement rock presents abundant planes of weakness due to the presence of phyllosilicates and to the layered organization of minerals. Natural water interconnectivity in rocks adjacent the borehole was noted in one 30ft-thick zone around the borehole. The basement rock porosity was found to be 5.6%, which is low. However, the basement rock is weaker than the sedimentary formations, which may be advantageous to induce fractures to stimulate flow during EGS operations.

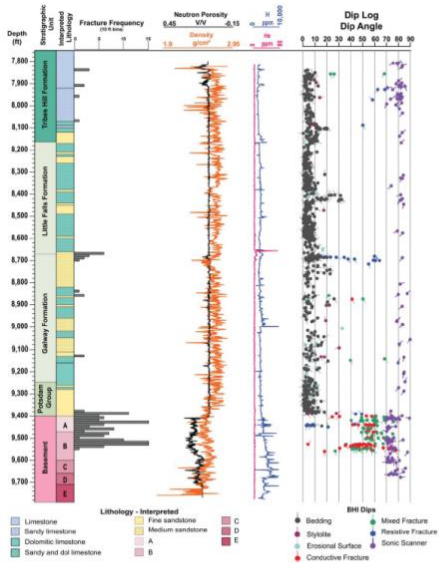
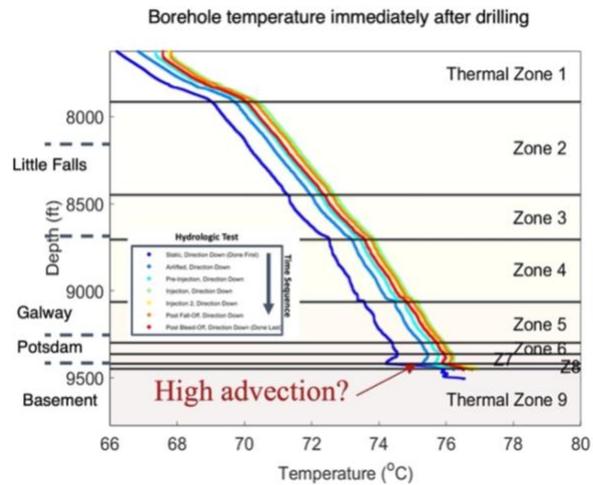

**Figure 11.** Thermal and hydraulic conditions in the lower 2000 ft of CUBO, where temperatures are sufficient for development of geothermal reservoirs. Source: Fulton et al., 2024, Stanford Geothermal Workshop.

Natural fractures that intersect CUBO were observed at three scales: from rock cores extracted from the walls of the borehole (observation length scale 0.001 – 400 mm), from micro-resistivity borehole image log data (observation length scale 1-25 cm) for which the fracture interpretation was guided by the overlapping core data, r, and sonic scanners whose sound wave propagation offers an indirect indication of open fractures (observation scales from 1-100 m). Overall, it was found that natural fractures are abundant in the sandstones of Galway and Potsdam, and that fracture scales range from microns to 100 m. The orientations of the fracture planes in the sedimentary rock are well determined. Small scale fractures in the sidewall cores are nearly all filled with mineral cement, but the few wider fractures display open space between mineral-lined walls. Fractures at greater apertures should transmit fluids. However, the actual apertures of the fractures detected by image logs or sonic data are unknown. It is thus important to determine whether the indirectly "observed" large fractures are consistent with the much higher confidence small fractures, and to understand whether the organization of the natural fracture sets at the scale of interest can serve as flow paths to use sedimentary rock as a geothermal reservoir. Three-dimensional statistical models of natural fractures in the sedimentary rocks based on high-confidence sidewall core data suggest that there are almost no connections among fractures and that the large, naturally interconnected fractures may contribute to reservoir flow only if they are further interconnected by stimulation. For the basement rocks, the natural fracture distribution is more uncertain, in part because criteria are lacking for distinguishing natural vs. drilling-induced fractures in the borehole image data. Fractures to a 100 m distance from borehole that contain fluids may be common, but they seem disorganized. Planes of natural weakness in the basement rock are abundant, but it is unclear whether interconnected flow path can be created.

**Cornell University Seismic Network (CorNet) (Patrick Fulton)**

The Cornell University Seismic Network (CorNet) is part of the International Federation of Seismographs Networks and has been operating since August 2019. It comprises 15 telemetered short-period instruments (seven buried at 30ft depth and eight buried 2ft deep). From January 2020 to 22 June 2023, 95 events were detected within 20 km of the proposed geothermal well site near CUBO, with 23 local magnitudes ranging from −1.02 to 0.56. The topography around the lake may create shallow stress concentrations along the lake axis, but none is expected beneath campus. Very small signals were measured around the noise level. This information is critical for characterizing the background microseismicity for comparison during future geothermal operations.

**Earth Source Heat Community Advisory Team (Sarah Carson, Marguerite Wells)**

The purpose of the Earth Source Heat Community Advisory Team (ESH CAT) was initially to develop a robust, long-term process of community participation to help prepare the community for the CUBO drill rig arrival Fall 2021, and to advise Cornell's decision-making processes on the feasibility of a subsequent ESH Demonstration Project. The ESH CAT still advises Cornell University on the steps towards a pilot EGS on campus. The main goals of ESH are to develop trust among community leaders and the Cornell team; respect community members' rights to have a voice in decisions that affect them; ensure that the community benefits from Earth Source Heat, create a sense of shared pride in project and climate leadership. ESH CAT has created early and frequent opportunities for interpersonal interaction between Cornell decision-makers and team members; connected with key opinion leaders in the community to ensure that they are and feel well-informed; provided proactive communication to inform the community on the CUBO activities, beyond what is required for permitting processes; provided mechanisms for community members to communicate questions, concerns, and feedback; solicited contributions from ESH CAT members to inform ESH messaging; and solicited ideas from ESH CAT members and identified stakeholder groups for community benefit opportunities. In total, ESH CAT has held eleven meetings since its inception and has engaged senior leaders in the conversations. During the construction of CUBO, in response to community feedback given to ESH CAT, the drill rig was switched to electric power to abate noise and reduce pollution concerns; a public viewing area was created, with routine "office hours" and an open house; and a web-based dashboard was created.

**Looking to the future: EGS prototype at Cornell (Wayne Bezner Kerr, Olaf Gustafson)**

A simple replacement of gas-fired heat sources with electric boilers to meet Cornell's 2023 heating requirements would result in a peak electric load of 103 MW, and an annual consumption of 311,000 MWhe. This peak load is more than three times larger than the current campus electric load (30 MW). Electrification solutions should maximize efficiency to minimize stress on regional electric grids and support grid reliability. The goal of the ESH team is to build a safe, reliable system capable of meeting campus heating loads for several years, which implies maximizing heat extraction and lateral length, minimizing thermal drawdown and capital expense, operating at low expense, and reducing the overall carbon footprint of the Ithaca campus, while collaborating in full transparency with the local stakeholders and communities. In the following, we discuss the design and build cycle of an EGS prototype made of an injection well and a production well at the vicinity of CUBO, as shown in Figure 12.a.

From a drilling energy point of view, the emission costs in $CO_2$ of a pair of geothermal wells with 6,000-ft laterals with fracture stimulation are equivalent to those of 1,422 shallow boreholes that could be used in a geoexchange system similar to Princeton's. In other words, deep EGS and shallow geoexchange technologies have similar carbon footprints if one only considers the drilling portion of the construction cycle. The calculation is as follows: each geoexchange well drilled on campus costs about 4,500 lbs of $CO_2$; the drilling of CUBO cost about 476,000 lbs of $CO_2$, which implies that a simple doublet with 6,000' laterals would cost about 2,000,000 lbs $CO_2$; stimulation of a pair of laterals is estimated at 4,400,000 lbs of $CO_2$. The construction of an EGS is less disruptive than that of a geoexchange installation, which requires drilling over large surface areas. Additionally, the monetary cost of simple EGS well pairs like those recently built in western United States is decreasing, in part thanks to knowledge gained from oil and gas completion technologies, and in part due to the lessons learnt from in commercial projects in Nevada and Utah. Challenges in deploying an EGS on the Ithaca campus include permitting complexity; subsurface uncertainties, particularly in terms of basement geology, reservoir temperature, and in situ stress; constraints on land ownership and boundaries; technological limitations; and cost. Over 30 permits, plans or required approvals have been identified. They impact every stage of planning and implementation, including drilling, traffic, stimulation, temporary power, noise, water use, and water disposal. Permits include processes at city, town, state, and federal levels. For example, a town site plan approval is needed, and requires permits for the complete demonstration system before drilling production wells. NYSDEC Division of Mineral Resources requires a drilling permit which is based on an environmental assessment and on design details on drilling, casing, cementing, blowout preventer, stimulation, and fluid management. Other permits cover erosion/stormwater, fill, trucking, construction trailers, waste disposal (cuttings, mud, water). Extra permits will be required, such as an injection permit from the Environment Protection Agency (EPA) for circulating geothermal water and New York State Public Service Commission (PSC) approval for extension of gas or electric lines (if required to power stimulation fleet).

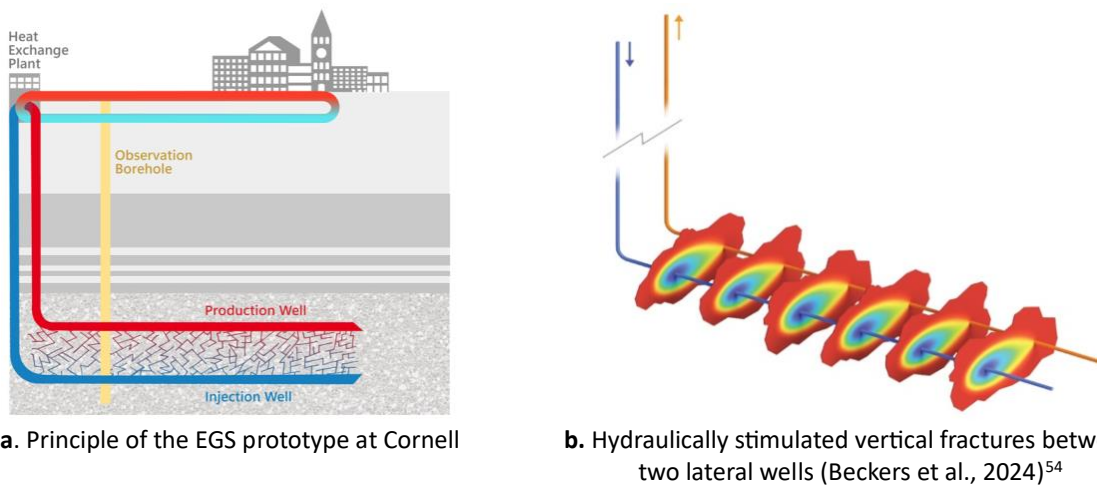

**a**. Principle of the EGS prototype at Cornell    **b.** Hydraulically stimulated vertical fractures between two lateral wells (Beckers et al., 2024)[54]

**Figure 12.** Principle of the EGS prototype envisioned on Cornell Ithaca campus: (a) injection, production and observation wells; (b) induced fracture pattern between the injection and production wells.

The site of the EGS is largely dependent on land ownership: Cornell property boundaries set the limits of the well plan. Additionally, drilling efforts planned for the EGS prototype on campus will be accompanied with the deepening of CUBO, such that the location of the well pad is constrained by the position of the CUBO well, since it would be very complicated to move the rig to a new well pad during the construction of the EGS. The orientation of in situ stresses sets the orientation of the lateral section of the wells: in order to propagate vertical fractures that can connect the injection and production wells as shown in Figure 12.b., the axis of the lateral wells must be aligned with the orientation of the minimal horizontal compression stress, i.e., NW-SE. A map of the tentative position and lateral orientation of the prototype EGS at Cornell is shown in Figure 13.a. The depth of the lateral wells ought to be large enough to avoid connecting the induced vertical fractures with gas bearing zones or with horizontal discontinuities between geological layers, which would lead to T-shaped fractures that short-circuit the circulation of the heat carrier. From a drilling perspective, the simplest well design is the one that involves the lowest rate of change of the direction of the well path, also called Dog Leg Severity (DLS), uses the least materials, and costs the least amount of money. The simplest design involves wells that change orientation from vertical to horizontal at rates as low as 1.5 degrees per 100 meters as shown in Figure 13.b. However, that geometry would limit the lateral portion of the well to less than 3,300-feet within the boundaries of the land owned by Cornell, which is unlikely to ensure the heat flow rate that is needed to demonstrate the feasibility of EGS heat direct use. A more complex deviated well plan could first deviate from vertical on a north-westerly heading, before building a curve through vertical and deviating in a direction opposite to the initial curve, such that the lateral runs on a south-easterly heading from the heel to the toe. A complex deviated well built with low DLS could be built with a lateral length greater than 6000' while remaining wholly within boundaries of Cornell property. In the design shown in Figure 13 c. the vertical observation well (blue) is advantageously positioned to allow observation of stimulation activity in the lateral sections of injection and production wells. The observation well will be completed with fiber optic cable permanently cemented outside the casing, along with pressure and temperature gauges. Stimulation of the lateral sections of producer and injector wells will be monitored via Distributed Acoustic Sensing (DAS), with improved sensitivity to strain associated with stimulation activity at all stages from the toe to the heel compared to observations from a vertical well located "behind" the heel as in Figure 13 b. The curvature of the well imposes limitations on the installation of the casing, the setting of which becomes more challenging with increasing DLS. Longer curved wells require more casing, which increases the weight that the rig will lift. These considerations reduce the flexibility of the design in terms of casing. The amount of energy needed to pump the reservoir to ensure a sufficient flow rate with the anticipated 7-inch casings is higher than with larger casings, due to increased pressure drop. Additionally, well stability concerns in complex wells may require setting additional casing strings, requiring bigger bits than low-DLS wells with simpler designs. Torque and Drag and Hydraulic limitations appear, and all materials and logistical challenges become greater.

---

[54] Beckers, K., Ketchum, A., & Augustine, C. (2024). *Evaluating Heat Extraction Performance of Closed-Loop Geothermal Systems with Thermally Conductive Enhancements in Conduction-Only Reservoirs* (No. NREL/CP-5700-88557). National Renewable Energy Laboratory (NREL), Golden, CO (United States).

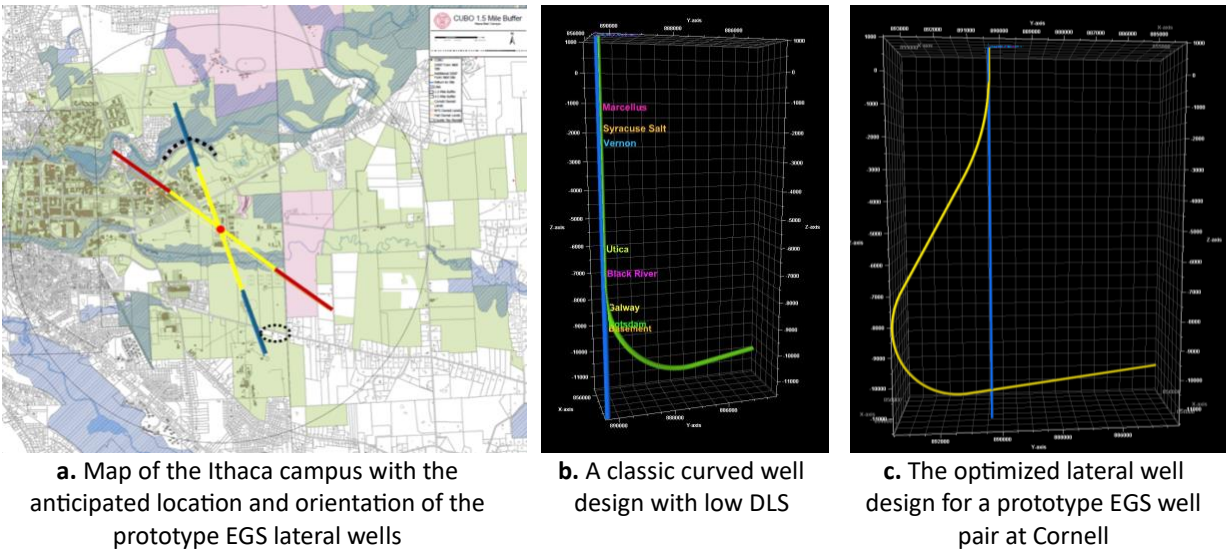

**a.** Map of the Ithaca campus with the anticipated location and orientation of the prototype EGS lateral wells

**b.** A classic curved well design with low DLS

**c.** The optimized lateral well design for a prototype EGS well pair at Cornell

**Figure 13.** Design of a prototype EGS on Cornell Ithaca campus: well heads location, lateral wells orientation, wells curvature.

From the thermal, hydraulic, mechanical and geological analyses performed at CUBO, it is now clear that stimulation will be needed to ensure the success of an EGS for direct heat use on the Ithaca campus of Cornell University. Stimulation plans must follow regulations and must be acceptable to community members. Stimulation is noisy and energy intensive, but short in duration. For one stage of plug-and-perf stimulation, 500,000 – 800,000 gallons of water and 500,000 – 600,000 lbs of sand are needed. The injected fluid may include acid treatment, biocides, viscosity modifiers, and/or friction reducer. Injection for stimulation requires 25MW of electricity, which is similar to the current campus load. The central energy plant on campus cannot provide the power necessary for stimulation. While diesel stimulation fleets are commonly used in the industry, it was already determined that the energy required for stimulation at Cornell would rely on gas- or electrical grid- powered generators. Several options are being explored to handle and dispose of the drilling and stimulation waste. The drilling mud will have to be hauled away, and so will the cuttings, unless some of them can be used as proppants during stimulation. The volume of stimulation and flowback water produced for the construction of a prototype EGS well pair is estimated at 18 M gals. Several options are being considered to treat wastewater onsite for local disposal, which would reduce the trucking demand. Social license will be crucial, particularly because permitting for chemicals, power, noise, and trucking is complex, and because quiet fleets and expert personnel are rare.

After drilling and stimulation, the ESH team anticipates running a flow test to assess the heat flow rate that can be produced from a well pair. From a technological standpoint, the easiest method to do so is to run a diesel pump through cascading tanks/ponds until the extracted fluid is cool enough to be re-injected. It may be necessary to use temporary cooling towers to impose a thermal load and/or new electric power on site to minimize emissions. Zoning, permits, and appeals may be required to get temporary structures. The operation will necessitate seismic monitoring, downhole fiber and pressure and temperature monitoring 24/7.

End-of-life closure and decommissioning for direct-use geothermal systems is minimal due to low surface footprint. NYSDEC regulations require that wells be properly closed ("plugged and abandoned") when taken out of service. The operator must file a Notice of Intention to Plug and Abandon, apply for a permit, and report on the plugging. The top and bottom of each casing string must be plugged with cement, and any open sections of well must be filled with stable fluid. Uncased freshwater zones must be sealed off with cement, the well head and related piping must be removed, and the well below grade must be permanently capped, after which, the site may be restored.

To summarize, building an EGS on Cornell Ithaca campus is much more expensive, challenging and complex than in other regions of the United States that target similar flow rates. EGS systems are relatively easy to close and reclaim, due to relatively small amounts of surface equipment. However, the path forward for EGS could be greatly eased by resolving questions of land use, mineral rights and access to heat. Permitting a project is enormously complicated and time consuming. State and federal regulators can encourage decarbonization by supporting regulatory harmonization.

## Conclusions

The ESH technology is fundamentally different than shale gas or oil production in that it removes stored thermal energy – it does not extract hydrocarbons, and heat extraction involves recirculation of water in the reservoir. The amount of stored thermal energy in the earth's crust that is accessible is massively large and is more ubiquitously distributed than fossil fuels. It is not limited to specific regions of the subsurface where hydrocarbons have accumulated. Equally important, ESH can restore itself, because heat conduction from the surrounding rock mass reheats locally cooled regions. The time scale for the reheating in conduction-dominated systems is about 3 to 4 times the heat extraction period and it would be faster when convection from regions outside the active reservoir is present. As a result, ESH can renew itself over a relatively short time which does not happen when hydrocarbon fluids are removed from shale gas or oil deposits. Utilizing geothermal energy directly for heating with ESH does not involve combustion of carbon contain compounds. The small footprint of ESH systems along with environmental benefits provided by avoiding carbon dioxide and fugitive methane emissions provide significant sustainability advantages by providing indigenous, affordable, clean energy for the long term.

Through research done at the Utah Frontier Observatory for Research in Geothermal Energy (FORGE), drilling bits were improved to sustain the harsh conditions of deep geothermal reservoirs, and drilling operations were optimized, leading to a Rate Of Penetration (ROP) of up 70ft/h. A variety of stimulation strategies were benchmarked, and a 90% recovery was achieved (hot fluid production rate of 24L/s for an injection rate at 27L/s). It was found that better efficiency is achieved when stimulation is done prior to drilling, and when proppants are used during stimulation. Lessons learnt at Utah FORGE are the most transferrable to the Cornell site. Indeed, other promising results were obtained at other sites, such as the Delft Subsurface Urban Energy Lab (DSUEL) and SuperHot Rock formations, but in geological conditions that are not comparable to Ithaca's. Data collected at Cornell University Borehole Observatory (CUBO) indicate that advective flow zones at temperatures amenable to direct heat production exist in the sedimentary rock and in the basement, but that none of them has sufficient permeability to operate an EGS without inducing fractures via stimulation techniques. The basement rock is easier to drill than the sedimentary rock, and temperature in the basement is higher than in the upper sedimentary layers. However, the presence of phyllosilicates in the basement makes it challenging to estimate the distribution of lengths and orientations of natural discontinuities. The orientation of the state of stress is known, and so are the magnitudes of the vertical stress and minimum compressive horizontal stress. But the estimation of the magnitude of the maximum horizontal compressive stress requires further verification. The current design envisioned for a prototype geothermal well pair at Cornell involves wells that present a curvature that does not vary monotonically, which presents important technological challenges. Recent progress made in computational geomechanics and Artificial Intelligence suggests that it will be possible to optimize drilling operations on-the-fly, simulate complex fracturing patterns for various stimulation scenarios and natural fracture distributions, and estimate the short- and long-term thermo-hydro-mechanical state of the geothermal reservoir. Funding, permitting and regulation remain important challenges ahead. Additionally, it is crucial to collaborate with various local stakeholders to obtain the social license to operate.

The ultimate goal of this workshop was to form an inter-disciplinary team that is prepared to write a proposal for an NSF Engineering Research Center (ERC) or a DOE Energy Frontier Research Center (EFRC) when a funding opportunity is announced in an area related to subsurface energy systems. During the workshop, participants demonstrated interest in sustained collaborations and discussed key aspects of related research endeavors, including: (1) Cornell's energy consumption in terms of end use, climatic constraints and production means available; (2) the potential impact of running an EGS facility on campus on Cornell's energy and water consumption, and on greenhouse gas emissions; (3) the full life cycle of EGS, including construction, exploitation, monitoring, maintenance, closure, reclamation and conversion. The workshop highlighted the extent to which EGS research and education is a multidisciplinary project that Cornell is uniquely well placed to undertake. By contrast with other technologies that engage only one or two colleges, deep geothermal energy presents institution-wide opportunities and interest. Indeed, Cornell workshop speakers were from the ESH project team, the School of Civil and Environmental Engineering, the Department of Earth and Atmospheric Sciences, the Smith School of Chemical and Biomolecular Engineering, the Department of Communication, Brooks School of Public Policy, and the School of Philosophy. In addition to those units, Cornell entities represented among participants included the Department of Biological and Environmental Engineering, the Department of Materials Science and Engineering, the Sibley School of Mechanical and Aerospace Engineering, and the Atkinson Center for Sustainability (Appendix 2). Just in 2024, teams of Cornell engineers and scholars led by Bert Bland (Cornell Associate Vice President for Energy and Sustainability) and Chloé Arson (Cornell Professor in Civil and Environmental Engineering) submitted a $17M DOE proposal[55], a $3M NSF preproposal for a

---

[55] Demonstration of Eastern US Enhanced Geothermal System (Cornell University), proposal submitted to the Department of Energy, EGS demonstrations. Total budget: $17M for 4 years, including $14.2M from DOE and $2.8M in cost-sharing. Lead PI: B. Bland. Co-PI: C. Arson. Aspiring PI: P. Fulton. Senior advisors: G. Abers, J. Tester.

Research Traineeship (NRT) program[56], and a response to a DOE Geothermal Energy from Oil + gas Demonstrated Engineering (GEODE) request for information (Appendix 3). The same team earned a Rapid Response Fund grant from the Atkinson Center for Sustainability[57] to organize this workshop, and the present report will serve as a basis to train starting graduate students in the workshop participants' respective laboratories, prepare white papers to funding agencies, and write a collective high-impact peer-reviewed publication. Workshop participants already identified future collaborative proposals, including another attempt to the NSF NRT program[58] and the NSF Regional Resilience Innovation Incubator (R2I2)[59].

---

[56] NRT-HDR: Data-driven equitable and sustainable subsurface energy engineering, preproposal submitted to the Office of the Vice President for Research and Innovation (OVPRI) for selection to the National Science Foundation Research Traineeship (NRT) program. Total budget: $3M for 5 years. Lead-PI: C. Arson. Co-PIs: M. Pritchard, S. Hormozi, S. Olmstead. Preproposal declined by the OVPR, to be resubmitted in 2025 after edits.
[57] Workshop: The role of Enhanced Geothermal Systems in the energy transition at Cornell, Cornell Atkinson Center for Sustainability, Rapid Response Fund, Grant 2024-RRF-Arson-cfa36, 2024. PI: C. Arson.
[58] https://new.nsf.gov/funding/opportunities/us-national-science-foundation-research-traineeship-program
[59] https://new.nsf.gov/funding/opportunities/r2i2-regional-resilience-innovation-incubator?utm_medium=email&utm_source=govdelivery

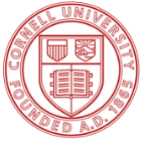

# Appendix 1: Workshop Agenda
Wednesday October 23 & Thursday October 24, 2024
Ithaca campus, Maple 120 132 Multipurpose Room

*This workshop is supported by Cornell Atkinson Center for Sustainability*
*(Grant 2024-RRF-Arson-cfa36)*

**Wednesday October 23**

**9am-9.30am** Welcome, purpose of the workshop (Chloé Arson)

**9.30am-10.30am** Cornell's climate goals (Sarah Carson, Stacey Edwards, Cole Tucker)
- Overview of Cornell's energy consumption, production and resources
- Review of University climate goals

**10.30am-11am** Coffee break

**11am-12pm** Ethics, justice and EGS
- Public Trust and Renewable Energy (Trystan Goetze)
- EGS regulatory framework vs. environmental justice (Sheila Olmstead)

**12pm-1pm** Lunch break

**1pm-2pm** EGS public acceptance (Sarah Carson, Katherine McComas, Marguerite Wells)
- State of the art on communication strategies to increase public acceptance
- Case study: EGS in NY State and the East of the United States

**2pm-2.30pm** Lessons learnt from the Cornell University Borehole Observatory (CUBO)
- Geology of the site, mechanical, thermal and hydraulic properties of the formation (Terry Jordan)

**2.30pm-3pm:** Coffee break

**3pm-4pm** Lessons learnt from DOE funded Frontier Observatory for Research in Geothermal Energy (FORGE) in Utah (Koenraad Beckers)
- Technological advances
- Feasibility of Enhanced Geothermal Systems (EGS) in Utah
- Fervo energy

**Thursday October 24**

**9am-10.30am** Current EGS research landscape (1h presentations followed by 30 min Q&A)
- The Cornell University Seismic Network (CorNet) (Patrick Fulton)
- Comparisons and contrasts with other academic geothermal drilling projects (FORGE, UrbEnLab, St1) (Patrick Fulton)
- Advanced geophysical monitoring: lessons learned from meso-scale EGS experiments (Seth Saltiel)
- 'Geothermal Everywhere' into the future: SuperHot Rock energy (Seth Saltiel)
- Novel computational geomechanics methods for EGS (Chloé Arson)

**10.30am-11am** Coffee break

**11am-12pm** Review of the EGS life cycle (Wayne Bezner Kerr, Bert Bland, Burak Erdinc, Olaf Gustafson)
- Site characterization and preparation including permitting
- Design
- Drilling, installation and monitoring
- Operation: Distribution and storage
- Facilities maintenance and waste management
- Closure, reclamation and conversion
- Site characterization and preparation including permitting

**12pm-1pm** Lunch break

**1pm-2.30pm** Panel: EGS impact metrics (Michael Gillenwater, Jeff Tester, Tony Ingraffea)
The moderator (Chloé Arson) will ask questions about environmental accounting at the energy systems level.

**2.30pm-3pm** Coffee break

**3pm-4pm** White paper preparation and wrap-up (Chloé Arson)
- White paper write-up:
  - Anticipated technological challenges and opportunities for EGS at Cornell
  - EGS impact metrics and LCA methodology
  - EGS impact and LCA at Cornell (various scenarios explored by the groups)
  - EGS ethics and public acceptance
  - Collaborative research ideas
- Wrap up:
  - Repository with group presentations and white paper draft
  - Dissemination plan for the white paper: arXiv, conference, journal?
  - Plans to follow up and collaborate for interested participants

**Expected deliverable:**
White paper and potential group publications on the topics discussed during the workshop.

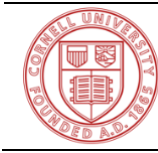

# Appendix 2: Workshop participants registered

| Cornell faculty members | | |
|---|---|---|
| Geoffrey Abers | Earth & Atmospheric Sciences | abers@cornell.edu |
| C. Lindsay Anderson | Biological & Environmental Eng. | cla28@cornell.edu |
| Chloé Arson | Civil & Environmental Eng. | cfa36@cornell.edu |
| Dominic Balog-Way | Department of Communication | db729@cornell.edu |
| Grace Barcheck | Earth & Atmospheric Sciences | cgb92@cornell.edu |
| Sriramya Duddukuri Nair | Civil & Environmental Eng. | sn599@cornell.edu |
| Nicole Fernandez | Earth & Atmospheric Sciences | nmf46@cornell.edu |
| Patrick Fulton | Earth & Atmospheric Sciences | pmf64@cornell.edu |
| Greeshma Gadikota | Civil & Environmental Eng. | gg464@cornell.edu |
| Trystan Goetze | School of Philosophy | tsg43@cornell.edu |
| Sarah Hormozi | Chemical & Biomolecular Eng. | sh2365@cornell.edu |
| Anthony Ingraffea | Civil & Environmental Eng. | ari1@cornell.edu |
| Teresa Jordan | Earth & Atmospheric Sciences | tej1@cornell.edu |
| Donald Koch | Chemical & Biomolecular Eng. | dlk15@cornell.edu |
| Katherine McComas | Department of Communication | kam19@cornell.edu |
| Gregory McLaskey | Civil & Environmental Eng. | gcm8@cornell.edu |
| Carolina Munoz-Saez | Earth & Atmospheric Sciences | cpm226@cornell.edu |
| Sheila Olmstead | School of Public Policy | smo74@cornell.edu |
| Seth Saltiel | Earth & Atmospheric Sciences | sas697@cornell.edu |
| Jefferson Tester | Chemical & Biomolecular Eng. | jwt54@cornell.edu |
| Uli Wiesner | Materials Science & Eng. | ubw1@cornell.edu |
| Alan Zehnder | Mechanical & Aerospace Eng. | atz2@cornell.edu |
| Max Zhang | Mechanical & Aerospace Eng. | kz33@cornell.edu |
| **Cornell energy and sustainability engineers and personnel** | | |
| Wayne Bezner Kerr | ESH Program Manager | wb264@cornell.edu |
| Robert Bland | As. VP, Energy & Sustainability | rrb2@cornell.edu |
| Sarah Carson | Director, Campus Sustainability | sc142@cornell.edu |
| Stacey Edwards | Energy Transition Program Manager | sae6@cornell.edu |
| Burak Erdinc | ESH Geothermal Engineer | ibe6@cornell.edu |
| Olaf Gustafson | Engineer Architect | jg72@cornell.edu |
| Cole Tucker | Director Utilities Distrib. & Energy Manag. | cmt233@cornell.edu |
| **Cornell Atkinson Center for Sustainability** | | |
| Shaun Doherty | Energy Transition & Carbon Management | sjd254@cornell.edu |
| **External to Cornell University** | | |
| Koenraad Beckers | NREL | Koenraad.Beckers@nrel.gov |
| Michael Gillenwater | GHG institute | michael.gillenwater@ghginstitute.org |
| Marguerite Wells | ACE NY | Mwells@aceny.org |

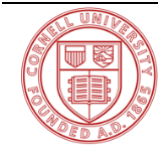

# Appendix 3: DOE GEODE Capabilities of Universities Request for Information – Response from Cornell University sent on 10/31/2024

**GENERAL**

1. How big is your group of staff researchers?
   - ~25 faculty members and > 50 postdocs and graduate students in Earth and Atmospheric Sciences, Civil and Environmental Engineering, Chemical and Biomolecular Engineering, Mechanical and Aerospace Engineering, Materials Science and Engineering, Biological and Environmental Engineering, Communications, Public Policy, Philosophy
   - ~12 full-time non-faculty Cornell employees who work on the Earth Source Heat (ESH) project with expertise and responsibility for the campus energy systems, renewable energy development, geology, drilling, community engagement, finance, and large infrastructure project management

2. Have you participated in geothermal research, development, and demonstration? If your answer is yes, please describe.
   - Earth Source Heat (ESH) at Cornell
     https://earthsourceheat.cornell.edu/
     Characterization of the geological, thermal and hydrological conditions of the subsurface in the North-East of the United States, engaged Community Advisory Team
   - DOE-funded Cornell University Borehole Observatory (CUBO):
     https://sustainablecampus.cornell.edu/living-laboratory/projects/cornell-university-borehole-observatory-cubo
   - Faculty at Cornell University have received funding through the Earthshot Enhanced Geothermal Shot initiative.

3. Does your group/team conduct experimental work and/or modeling work related to geothermal engineering and technologies?
   - Back-calculation of rock in situ stress from measurements at Cornell University Borehole Observatory (CUBO)
   - CorNet: Cornell seismic monitoring network
   - Experimental and modeling work on active tracers to enhance the efficiency of the geothermal reservoir
   - Experimental design of cements for Enhanced Geothermal wells
   - Numerical modeling of pressurized fracture propagation

4. Have you worked in Oil and Gas? If yes, please describe.
   - Drilling
   - Oil and gas cement wells
   - Reservoir modeling
   - Simulation of hydraulic fracturing

5. Have you worked in high temperature, or other conditions specific to geothermal applications? If your answer is yes, please elaborate.
   - Geothermal wells at 200°C in Turkey
   - Cornell University Borehole Observatory (CUBO) at 80°C
   - Design of an Enhanced Geothermal System (EGS) at 80°C at Cornell
   - Low cost alternative cementitious materials for EGS in 150-300 C
   - Supplementary cementitious materials for EGS in 120-150 C.

6. Have you worked in subsurface characterization that can be applied to geothermal engineering and technologies? If your answer is yes, please describe.
   - CorNet: Cornell seismic monitoring network
   - During the 2022 CUBO project, a suite a of logging tools were run, including gamma ray, caliper, resistivity, density, neutron porosity, spectral gamma, micro resistivity imaging (FMI), sonic velocity, ultrasonic imaging and cement bond logging. Additional testing for subsurface characterization included mini-frac via MDT tool, side wall coring and P/T logging while flow testing via air lifting of well bore fluids and subsequent reinjection and pressurization. Data from all logging and testing has been analyzed internally by the Cornell team with assistance by external consultants and service providers where appropriate. Complete logging data, as well as mud logging results and active seismic data have been integrated into models internally, using industry software (Petrel, Techlog).

7. Of the below research areas, which does your department have **active** projects in that apply to geothermal or oil and gas technologies? If not applicable, please list N/A.
   a. Drilling Technology: N/A
   b. Completions and Stimulation: Proven track record
      Numerical modeling
      Experimental work
   c. Data Interpretation and Modeling: Proven Track Record
      Seismic monitoring network
      Instrumented observatory well
      Signal processing
   d. Rigs and Equipment: N/A
   e. Production and Operations: Proven Track Record
      High-temperature resistant cements
      Smart tracers
   f. Other that you consider useful for geothermal: Emerging Interest
      Numerical modeling of serpentinization for geological hydrogen extraction

**FACILITIES**

8. Which of the following areas does your group/team have facilities in? Please select all that apply and provide a brief description of each applicable facility. Where applicable, note whether the facility is laboratory-based (in a science building on a university campus) or a field laboratory (property owned by a university that is not necessarily on campus but where larger scale experiments can be conducted).

- Drilling Technology: N/A
- Completions and Stimulation
  Rock mechanics testing equipment (laboratory-based)
  HPHT cementing equipment (laboratory-based)
- Data Interpretation and Modeling - please list the computing power of your facility
  Cornell University Borehole Observatory (CUBO): instrumented borehole (field-based)
  Cornell seismic monitoring network (CorNet): field-based
  Reservoir modeling capabilities – multi-core workstations
  Fracture modeling capabilities – multi-core workstations
  The Cornell University Center for Advanced Computing (CAC) offers cloud computing, allowing uses to create instances (virtual machines) with up to 128 CPU cores and 240GB RAM that deploy in seconds. NVIDIA T4, V100, and A100 GPU instances are available. A Ceph cluster with 1.9PB of raw capacity provides the storage capabilities.
- Rigs and Equipment: N/A
- Production and Operations
  Smart tracer testing facility: laboratory-based
- Other that you consider useful for geothermal: N/A

**COMPETENCIES**

9. What research areas and projects are your group/team working on in 2025-2026 that apply to geothermal or oil and gas technologies?
   - Detection of advection zones
   - Induced seismicity
   - Super Hot Rock
   - Reactive transport in geothermal reservoirs
   - Thermally active tracers
   - Well cements resistant to high temperatures
   - Numerical modeling of pressurized fractures
   - Numerical modeling of complex geothermal well systems
   - Economical models of EGS
   - Sociological impacts of EGS
   - Environmental justice
   - Public trust in EGS

10. What are the geothermal energy-related emerging interests of researchers in your group/team that you are envisioning in 2025-2026? Please elaborate in a short paragraph on how these interests will be applied to geothermal technologies.
    - Seismicity inverse problems – from monitoring data to earthquake characterization
    - Super critical geothermal systems
    - Smart tracers that act as mobile sensors in geothermal reservoirs
    - Design and modeling of geothermal well cements
    - Multi-scale numerical models of pressurized fractures
    - Earth model digital twins for lateral well optimization and stress, pore pressure and temperature estimation

- Multiscale data integration for optimal geothermal reservoir characterization and modeling
- Augmented reality (A/R) in rock mechanics and geology for enhanced geothermal reservoir characterization
- Public policy for EGS development
- Best practices for EGS developers – community collaborations and trust

11. What is the geothermal energy-related expertise (as defined in the 'Definitions' section in the Introduction) of researchers in your group/team? If they are not directly related to existing geothermal technologies, please elaborate on how this expertise can apply to geothermal technologies in 2025-2026.
    - Drilling Technology
      Engineers in the ESH team at Cornell have extensive experience drilling oil and gas and geothermal wells.
    - Completions and Stimulation
      Engineers in the ESH team at Cornell have experience with several stimulation techniques, for both oil and gas and geothermal applications.
      Development of field deployable cement slurries.
    - Data Interpretation and Modeling, including AI and machine learning
      Engineers and faculty members at Cornell have extensive experience with the interpretation of large geological, hydrological, thermal, mechanical and seismic data sets. Faculty members and researchers develop forward numerical models (e.g., based on the Finite Element Method or the Finite Volume Method) to simulate reactive transport, hydraulic fracturing, and thermos-hydro-mechanical coupled processes in reservoirs. Faculty members and researchers also develop an array of machine-learning tools to interpret large data sets, accelerate numerical simulations and detect geomechanical patterns in field data or numerical results.
    - Rigs and Equipment
      Engineers in the ESH team have expertise in most of the equipment needed to build an EGS, including the rigs.
    - Production and Operations
      Engineers in the ESH team have expertise in hydraulic fracturing and other stimulation techniques. Research groups at Cornell are interested in optimizing EGs operations via forward numerical modeling and deep learning.
    - Other that you consider useful for geothermal: N/A

12. What research areas and projects are your group/team working on in 2025-2026 that apply to reservoir engineering, if any?
    - Reservoir characterization and fracture simulation for oil and gas applications
    - Carbon Capture and Storage (CCS)
    - Low-temperature Enhanced Geothermal Systems (EGS) for heat direct use
    - Super Hot Rock
    - Cements for EGS
    - Smart tracers for EGS
    - Geological hydrogen

13. What research areas and projects are your group/team working on in 2025-2026 that apply to geology, geochemistry, and/or geophysics, if any?
    - Reservoir characterization and fracture simulation for oil and gas applications
    - Carbon Capture and Storage (CCS)
    - Low-temperature Enhanced Geothermal Systems (EGS) for heat direct use
    - Super Hot Rock
    - Cements for EGS
    - Smart tracers for EGS
    - Geological hydrogen

**TRACK RECORD**

14. Please provide the ten most relevant papers, presentations, and patents that your group/team has published that pertain to geothermal technologies or oil and gas technologies that can apply to geothermal (provide brief details how those Oil and Gas technologies are relevant to geothermal technologies).

    a. Valentino, D.W., Chiarenzelli, J.R., Jordan, T.E., Jacobi, R.D., and Gates, A. E., 2024, Deep borehole discoveries beneath the Appalachian Basin: broad Rodinian rift and Neoproterozoic tectothermal event: *Terra Nova*, v. 36, p. 1-7, https://doi.org/10.1111/ter.12741

    b. Katz, Zachary S., Abers, Geoffery A., Yang, Yucheng, Ferris, Aaron, Jordan, Teresa E., Pritchard, Matthew E., Fulton, Patrick M., and Gustafson, Olaf, 2024, Seismic Monitoring near Ithaca, New York, Reveals Nonuniform Distribution of Microseismicity in an Intraplate Region. *Seismological Research Letters*: v. 95, doi: https://doi.org/10.1785/0220240158

    c. Beckers, K.F., Rangel-Jurado, N., Chandrasekar, H., Hawkins, A.J., Fulton, P.M., and Tester, J.W., 2022, Techno-Economic Performance of Closed-Loop Geothermal Systems for Heat Production and Electricity Generation: *Geothermics*, v. 100, p. 102318, https://doi.org/10.1016/j.geothermics.2021.102318

    d. Bergen, S. L., Zemberekci, L., & Nair, S. D., 2022. A review of conventional and alternative cementitious materials for geothermal wells. *Renewable and Sustainable Energy Reviews*, *161*, 112347, https://doi.org/10.1016/j.rser.2022.112347

    e. Beentjes, I., Bender, J.T., Hawkins, A.J., and Tester, J.W., 2020, Chemical dissolution drilling of barre granite using a sodium hydroxide enhanced supercritical water jet: *Rock Mechanics and Rock Engineering*, v. 53, p. 483–496. https://doi.org/10.1007/s00603-019-01912-7

    f. Jin, W., & Arson, C., 2020. Fluid-driven transition from damage to fracture in anisotropic porous media: a multi-scale XFEM approach. *Acta Geotechnica*, 15, 113-144, https://doi.org/10.1007/s11440-019-00813-x

    g. Beckers, K.F., Koch, D.L., and Tester, J.W., 2015, Slender-body theory for transient heat conduction: theoretical basis, numerical implementation and case studies:

*Proceedings of the Royal Society A: Mathematical, Physical and Engineering Sciences*, v. 471, p. 20150494, https://doi.org/10.1098/rspa.2015.0494

h. Beckers, K.F., Lukawski, M.Z., Anderson, B.J., Moore, M.C., and Tester, J.W., 2014, Levelized costs of electricity and direct-use heat from Enhanced Geothermal Systems: *Journal of Renewable and Sustainable Energy*, v. 6, p. 013141, https://doi.org/10.1063/1.4865575

i. Hwang CG, Ingraffea AR., 2007. Virtual crack extension method for calculating the second order derivatives of energy release rates for multiply cracked systems. *Eng. Fract. Mech.*, 74:1468-1487, https://doi.org/10.1016/j.engfracmech.2006.08.009

j. Boone TJ, Ingraffea AR., 1990. A Numerical Procedure for Simulation of Hydraulically - Driven Fracture Propagation in Poroelastic Media. *Int. J. Num. Analyt. Meth. Geomech.*, 14, 27-47, https://doi.org/10.1002/nag.1610140103

15. Please provide other relevant activities (such as training and developed curricula) that your group/team has access to that you think pertain to geothermal or oil and gas technologies.
    - NSF funded Integrative Graduate Education and Research Training program (IGERT), 2010-2019
    - Drilling scientific conference held at Cornell University, 2019
    - Drilling Well On Paper (DWOP) prior to the drilling of Cornel University Borehole Observatory (CUBO), 2021-2022
    - NSF Research Traineeship pre-proposal submitted internally at Cornell, June 2024
    - Workshop on the role of EGS in the energy transition at Cornell, October 2024, funded by the Cornell Atkinson Center for Sustainability

16. Please provide other technologies that your group/team has access to that you think may pertain to geothermal or oil and gas technologies. If not applicable, please list N/A.
    - Rock mechanics testing equipment (laboratory-based)
    - Smart tracer testing facility: laboratory-based
    - Cornell University Borehole Observatory (CUBO): instrumented borehole (field-based)
    - Cornell seismic monitoring network (CorNet): field-based
    - Reservoir modeling capabilities – multi-core workstations
    - Fracture modeling capabilities – multi-core workstations
    - The Cornell University Center for Advanced Computing (CAC) offers cloud computing, allowing uses to create instances (virtual machines) with up to 128 CPU cores and 240GB RAM that deploy in seconds. NVIDIA T4, V100, and A100 GPU instances are available. A Ceph cluster with 1.9PB of raw capacity provides the storage capabilities.

**ADDITIONAL COMMENTS**

17. Please provide any other information or data that are relevant to geothermal technologies that you would like us to consider.
    Stratigraphic analysis based on a borehole observatory on campus
    On-going research on Earth Source Heat (for EGS)